\documentclass[aps,superscriptaddress,prb,12pt]{revtex4-1}
\usepackage[utf8]{inputenc}
\usepackage{natbib}
\usepackage{graphicx}
\usepackage{dcolumn}
\usepackage{bm}
\usepackage{epsfig}
\usepackage{textcomp}
\usepackage{gensymb}
\usepackage{xcolor}
\newcommand{\new}[1]{{\color{black}{#1}}}

\begin{document}

\title[]{Selecting ``Convenient Observers" to Probe the Atomic Structure of Epitaxial Graphene Grown on Ir(111) via Photoelectron Diffraction}

\author{Lucas Barreto}
\affiliation{Centro de Ci\^{e}ncias Naturais e Humanas, Universidade Federal do ABC, Santo André 09210-580, SP, Brazil}

\author{Luis Henrique de Lima}
\affiliation{Centro de Ci\^{e}ncias Naturais e Humanas, Universidade Federal do ABC, Santo André 09210-580, SP, Brazil}

\author{Daniel Coutinho Martins}
\affiliation{Centro de Ci\^{e}ncias Naturais e Humanas, Universidade Federal do ABC, Santo André 09210-580, SP, Brazil}

\author{Caio Silva}
\affiliation{Instituto de F\'{i}sica Gleb Wataghin, Universidade Estadual de Campinas, Campinas 13083-859, SP, Brazil}

\author{Rodrigo Cezar de Campos Ferreira}
\affiliation{Instituto de F\'{i}sica Gleb Wataghin, Universidade Estadual de Campinas, Campinas 13083-859, SP, Brazil}

\author{Richard Landers}
\affiliation{Instituto de F\'{i}sica Gleb Wataghin, Universidade Estadual de Campinas, Campinas 13083-859, SP, Brazil}

\author{Abner de Siervo}
\affiliation{Instituto de F\'{i}sica Gleb Wataghin, Universidade Estadual de Campinas, Campinas 13083-859, SP, Brazil}

\begin{abstract}
Epitaxial graphene grown on metallic substrates presents, in several cases, a \mbox{long-range} periodic structure due to a lattice mismatch between the graphene and the substrate. For instance, graphene grown on Ir(111), displays a corrugated supercell with distinct adsorption sites due to a variation of its local electronic structure. This type of surface reconstruction represents a challenging problem for a detailed atomic surface structure determination for experimental and theoretical techniques. In this work, we revisited the surface structure determination of graphene on Ir(111) by using the unique advantage of surface and chemical selectivity of synchrotron-based  photoelectron diffraction. We take advantage of the Ir 4f photoemission surface state and use its diffraction signal as a probe to investigate the atomic arrangement of the graphene topping layer. We determine the average height and the overall corrugation of the graphene layer, which are respectively equal to  \mbox{3.40 $\pm$ 0.11 {\AA}} and \mbox{0.45 $\pm$ 0.03 {\AA}}. Furthermore, we explore the graphene topography in the vicinity of its high-symmetry adsorption sites and show that the experimental data can be described by three reduced systems simplifying the Moir\'{e} supercell multiple scattering analysis. 
\end{abstract}

\maketitle
\section{Introduction}
Since the isolation of a single graphene layer \cite{Novoselov2004}, a considerable effort has been performed to develop and modify fabrication processes to improve its quality \cite{Lee2019}. In this context, the synthesis of graphene on single-crystal metallic surfaces via chemical vapor deposition is a well-established route to obtain high-quality large-area graphene layers \cite{Wintterlin2009, Batzill2012, Lizzit2012}. Particularly, Ir(111) is a notable substrate to grow graphene. First, under certain controllable conditions, it is possible to systematically obtain a full single layer without rotational domains \cite{Hattab2011}. Second, the graphene-substrate interaction is relatively weak such that the graphene $\pi$-band is almost intact\cite{Pletikosic2009,kralj2011}. Actually, the electronic structure, which presents mini gaps and replica bands, can be tuned and chemically decoupled from the substrate by intercalation\cite{Larciprete2012, Ulstrup2014, Pervan2017, Balog2019}.

From the structural point of view, graphene grown on iridium (Gr/Ir) presents a lattice mismatch between the carbon layer and the underlying substrate, which forms a moir\'{e} superstructure \cite{Dedkov2015}. The large periodicity of the moir\'{e}  superlattice is reflected on a variation of the carbon substrate distance through the supercell such that the graphene layer displays a corrugated feature leading to a modulation of the electronic density \cite{Voloshina2013}. Therefore, the adsorption behavior along the moir\'{e} cell can vary significantly. In this sense, such moir\'{e} structures can be used as a template to form highly ordered dispersed \new{nanostructures}. For instance,  Gr/Ir has been used to create periodic ordered metallic nanoclusters \cite{Ndiaye2006} and atomic periodic superlattices \cite{Baltic2016, Petrovic2017, Pivetta2018}. 

The interest in Moir\'{e} structures go beyond graphene on metals. For example, transition metal dichalcogenides such as MoS$_{2}$ and WSe$_{2}$ both grown on Au(111) exhibit a \mbox{corrugated} Moir\'{e} structure \cite{Sorensen2014, Gronborg2015, Dendzik2015}, also hexagonal boron nitride (h-BN) grown on several metallic substrates shows similar features \cite{Auwarter2019}.  Moreover, the weak interaction between layers in \mbox{two-dimensional} materials opens the possibility of stacking atomic layers with different lattice parameters forming a Moir\'{e} superlattice. One can also form a long-range ordered structure twisting two layers of the same material \cite{Cheng2019}. In those structures, the electrons are under the influence of a long-range potential, which can lead to emergent phenomena such as superconductivity and formation of flat bands \cite{Cao2018, Zhao2020}.     
    
The position of the carbon atoms on the corrugated graphene layer is directly connected to its local electronic structure. Therefore, determining the surface topography is \mbox{crucial} to understand the behavior of the different adsorption sites. The modulated surface can be \mbox{directly} observed by imaging techniques such as Atomic Force Microscopy (AFM) \cite{Hamalainen2013} and Scanning Tunnelling Microscopy (STM) \cite{Diaye2008, Ferreira2018}. Furthermore, the modulation can also be identified via Low Energy Electron Diffraction (LEED), where satellites surround the main reflection spots of the substrate \cite{Hattab2011}.  

In order to determine the distance between the carbon layer and the substrate, it is necessary to use scattering techniques  \cite{Busse2011, Jean2015}. Although LEED is mostly used to examine the quality of graphene grown on metallic substrates \cite{Hattab2011, Rogge2015}, the quantitative LEED analysis is challenging since one has to deal with dynamical scattering calculations on a considerably large unit cell. For instance, in the case of Gr/Ir and Gr/Ru, the typical unit cells consider hundreds of scattering centers\cite{Moritz2010, Hamalainen2013}. In fact, a proper structural characterization of those long-range superstructures demands novel description methodologies \cite{Hermann2012,Zeller2014,Moritz2015,Zeller2017,Kuznetsov2018,Ster2019, Lima2020}.

In this work, we investigate the Gr/Ir atomic structure via synchrotron-based X-ray Photoelectron Diffraction (XPD). \new{We use the substrate diffraction signal to probe the graphene structure\cite{Muntwiler2001}. Notably, we have selected the photoelectrons from the topmost substrate layer due to the presence of the iridium surface state. This approach makes the XPD signal much more sensitive to the C layer, avoiding the strong forward scattering on Ir layers observed for the Ir bulk signal. Therefore,}  the unique chemical selectivity of the technique combined with the chance of selecting the photon energy allows us to increase the \new{probability of having scattering events at the carbon layer}.  Instead of performing the XPD multiple scattering analysis on the entire supercell, we reduce the Gr/Ir Moir\'{e} superstructure into three reduced clusters. \new{Based on the location of the high-symmetry sites, we select one electron emitter, a \textit{convenient observer}, on each cluster to probe the graphene topography around the selected emitter. Although this approach does not consider the problem  extensively, it dramatically reduces the computation effort for the complex surface structure determination allowing us to describe the corrugated Gr/Ir atomic structure with an acceptable agreement between theory and experiment.}    

\section{Methods}
\begin{figure}[!htb]
\centering
\includegraphics[scale=0.25]{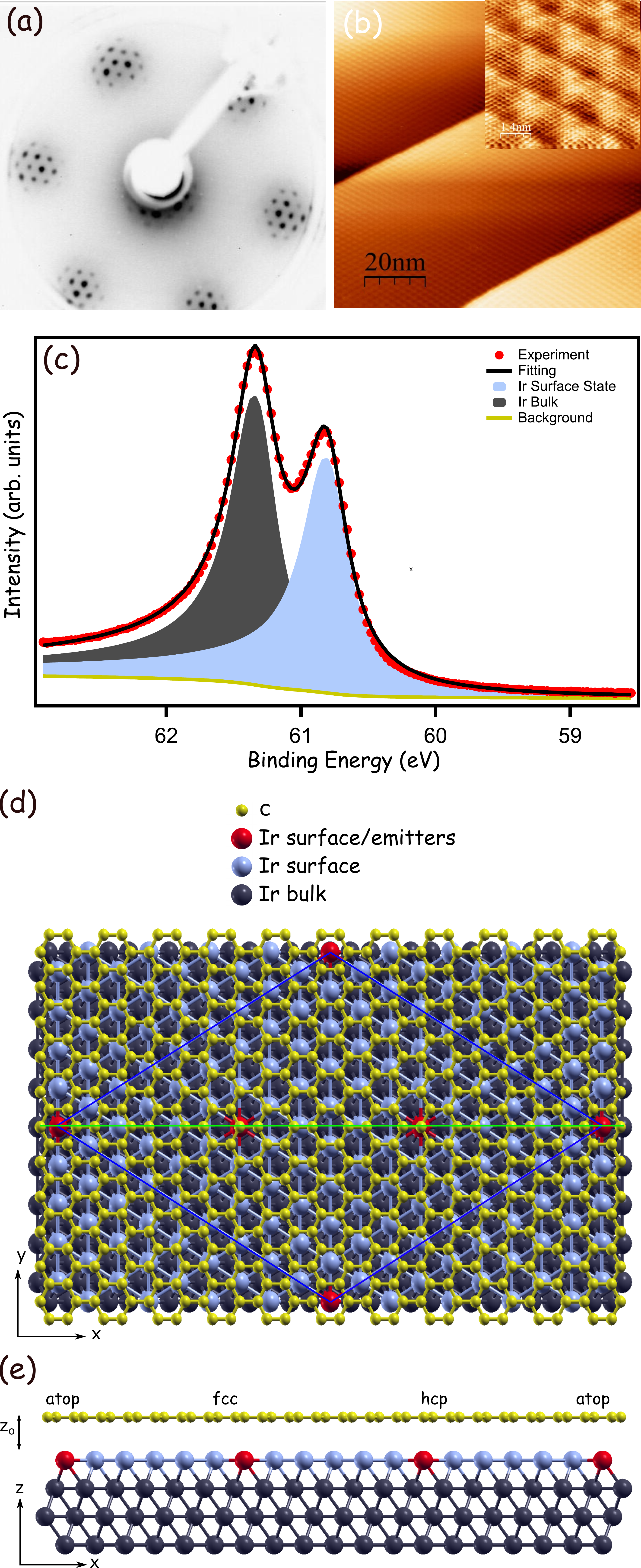}
\caption{Gr/Ir characterized by (a) LEED (E=$70$ eV), (b) STM (100 nm $\times$ 100 nm, U$_{bias}=0.6$ V, I$_t=0.57$ nA), inset (7 nm $\times$ 7 nm, U$_{bias}=0.13$ V, I$_t=0.57$ nA), and (c) HR-XPS (h$\nu=190$ eV). In (c), it is also shown the Ir 4f surface and bulk components obtained by peak deconvolution. (d) Top   and (d) side view of the Moir\'{e} superstructure model. The side view is a cut through the green line shown in (d). \new{In this case, the cell is considered to be a $10 \times 10$ graphene layer on a $9 \times 9$ Ir(111) substrate ($a=24.4$ \AA).}}
\label{fig1}
\end{figure}

The Ir(111) crystal was cleaned by sputtering using 1.0 KeV Ar ions followed by flash annealing up to 1570 K. A full graphene layer was obtained via thermal decomposition of hydrocarbons: the sample was kept at 1570 K in a $5 \times 10^{-7}$ mbar propylene background pressure. From the LEED pattern exhibited in the inset figure \ref{fig1}a, it is possible to conclude that the graphene does not present rotational domains \cite{Hattab2011, Rogge2015}. Moreover, the STM images in figure \ref{fig1}b reinforce the high quality of the graphene layer since a low defect density is observed. \new{For the structural determination, we performed angle scan synchrotron-based  XPD measurements combined with multiple scattering analysis using the Multiple Scattering Calculation of Diffraction (MSCD) package \cite{Rehr1990,Chen1998}. Based on an atomic cluster, we calculate the theoretical diffraction pattern and compare it with the experimental data through a reliability factor ($R_a$) defined in such a way that $R_a=0$ indicates a perfect agreement between theory and experiment \cite{Siervo2002, Soares2002}. Thus a set of parameters, in a trial and error approach, is varied to minimize $R_a$. See supplemental material for a detailed description of the XPD experiment\cite{Cezar2013}, structural determination, and error analysis \cite{Booth1997,Pendry1980,Bondino2002}.}

\section{Discussion and Results}
For the clean Ir(111) surface, the Ir 4f$_{7/2}$ XPS spectrum exhibits two components: the lower binding energy peak is a surface state (SS) associated with the topmost iridium atomic layer, and the other component is related to the underlying atoms \cite{Veen1980, Bianchi2009, Ferreira2018}. Figure \ref{fig1}c shows a high-resolution XPS of Gr/Ir where it is possible to observe the iridium SS, which is not affected by the presence of the carbon layer \cite{Lacovig2009, Larciprete2012, Ferreira2018}. Therefore, for the structural analysis, we use the diffraction signal originated from the SS peak to narrow down the number of emitters and concentrate on the first iridium atomic layer, represented by red and blue in figures \ref{fig1}d-e. However, even in this scenario, the number of emitters is considerably large for a typical XPD multiple scattering calculation. For this reason, in the simulations, we only consider three emitters in the entire unit cell, which are shown in red in figures \ref{fig1}d-e. The selection of those specific emitters is related to their proximity to the high-symmetry graphene adsorption sites: atop, fcc and hcp.      

\begin{figure*}[!htb]
\centering
\includegraphics[width=\textwidth]{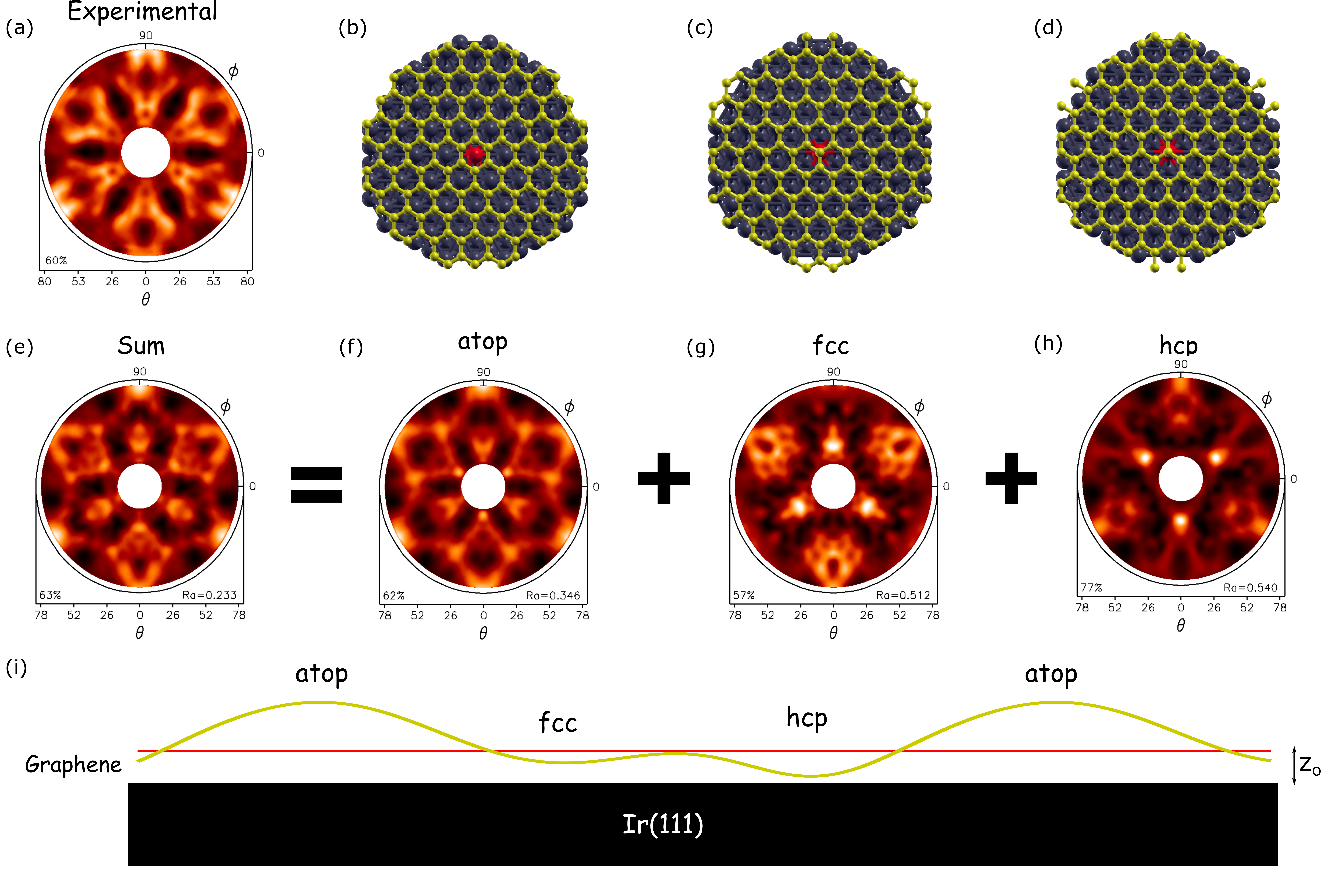}
\caption{(a) Ir SS experimental XPD pattern ($h\nu=122$ eV). (b)-(d) Top view of the atop, fcc and hcp clusters used in the  simulations. (e) Best theoretical XPD pattern which is obtained summing the patterns associated with the (f) atop, (g) fcc, and (h) hcp clusters. The percentage displayed on the bottom-left of each XPD pattern is its maximum anisotropy \cite{Westphal2003}. (i) Resulting graphene layer profile (not in scale).}
\label{fig2}
\end{figure*}

The experimental diffraction pattern obtained from the Ir 4f$_{7/2}$ SS is exhibited in figure \ref{fig2}a. Notably, the kinetic energy of the photoelectrons is $\sim 62$  eV, which leads to an inelastic mean free path lower than $5$ \AA\  on iridium \cite{Tanuma1991,Powell2010}. \new{Therefore, the possibility of selecting the Ir SS photoelectrons combined with their low kinetic energy increases the surface sensitivity. The former reduces the forward focusing contributions of the inner Ir layers present in the Ir 4f bulk signal. The latter decreases the probability of electrons that eventually travel into deep layers  of the substrate through backscattering reaching the detector.} Additionally, the distance between the selected emitters in the supercell is $\sim 14$ \AA\  such that their contributions to the diffraction pattern can be isolated or, put differently, their contribution to the total intensity can be summed incoherently. On that premise, we split the supercell into three clusters with a semi ellipsoidal shape \cite{Chen1998} centered on each emitter. Figures \ref{fig2}b, c, and d show, respectively, the top view of the clusters associated to the atop, fcc and hcp sites; each of them has more than 370 atoms and a surface radius equal to 11 \AA. 
  
The graphene-iridium distance ($d_{C-Ir}$) variation through the unit cell leads to a large number of structural parameters to be determined. Since it is possible to observe a smooth graphene topography via scanning probe techniques such as AFM and STM \cite{Hamalainen2013, Ulstrup2014, Ferreira2018}, we assume that the graphene topography can be described by continuous functions, particularly Gaussian functions \cite{Lima2013,Ulstrup2018}. Similarly, in some structural studies, $d_{(C-Ir)}$ is expressed by a truncated Fourier series \cite{Busse2011, Hamalainen2013, Jean2015}. In contrast, for h-BN grown on Rh(111), the height transition along the supercell is so abrupt that the h-BN topography can be modeled by distinct flat areas \cite{Lima2020}. Here, the graphene-substrate separation at an in-plane location $(x,y)$ (see figure \ref{fig1}d) on each cluster $i$ is described by:   
\begin{equation}\label{eq1}
d_{(C-Ir)_{i}}=z_0+ A_{i} \mathrm{e}^\frac{{-\left[\left(x-x_{i}\right)^{2}+\left(y-y_{i}\right)^{2}\right]}}{B_{i}}, i=fcc,hcp,atop,
\end{equation}
\noindent where $A_{i}$ and $B_{i}$ are, respectively, the parameters related to the Gaussian amplitude and width,  the pair ($x_{i}$,$y_{i}$) is the location of the emitter $i$ and $z_{o}$ is a distance offset, see figure \ref{fig1}e. The index $i$ in equation \ref{eq1} is related to the three considered emitters, which are termed based on their proximity to the graphene site. For example, the atop emitter is located under the atop graphene adsorption site. The same procedure is used to label the other emitters. 

The diffraction pattern which provides the best agreement between theory and experiment is shown in figure \ref{fig2}e from which we conclude that the proposed model provides a very good description of the experimental data ($R_{a}=0.23$). The diffraction pattern of each cluster is exhibited in figures \ref{fig2}f-h; an isolated emitter provides an inadequate description of the experimental data, as can be seen by the $R_{a}$ values. The fitting parameters for the best structure are summarized in table \ref{table1}. For the fcc and hcp sites, the  Gaussian prefactor is negative, indicating that in the vicinity of those emitters, the carbon atoms are closer to the substrate. Inversely, in the atop site, the graphene layer is located at a higher distance from the substrate. We also investigated the presence of  buckling in the graphene layer, but, differently from graphene on SiC \cite{Lima2013,Lima2014,Lima2014a} and black phosphorus \cite{Lima2016}, we could not identify any significant vertical displacement between the surface sublattices. A profile along the green line of figure \ref{fig1}d is displayed in figure \ref{fig2}i. The graphene layer is closer to the substrate in the hcp site in agreement with previous structural analysis \cite{Busse2011, Hamalainen2013}.     

\begin{table}[!htb]
   \begin{center}
    \begin{tabular}{c|c|c|c}
                     &        fcc     &    hcp          & top                  \\ \hline \hline    
   A$_{i}$ (\AA)           & -0.14 $\pm$ 0.02   &   -0.19 $\pm$ 0.02   &  0.26 $\pm$ 0.02             \\ \hline  
   B$_{i}$ (\AA$^{2}$)     &  65   $\pm$  12  &     38  $\pm$ 10  &  90   $\pm$  11  \\ \hline 
   z$_o$ (\AA)           & \multicolumn{3}{c}{3.39 $\pm$ 0.03}                         \\ \hline  
     \end{tabular}
    \caption{Fitting parameters for the structure which presents the best theory-experiment agreement. The parameters $A_{i}$, $B_{i}$ and $z_{o}$ are defined by equation \ref{eq1}.}
    \label{table1}
  \end{center}
\end{table} 
    
The overall corrugation ($\Delta \bar{d}_{C-Ir}$) is evaluated as the difference between the highest and lowest graphene distance from the substrate. Furthermore, the coordinates of the carbon atoms on each cluster are used to estimate the graphene mean height ($\bar{d}_{C-Ir}$). The obtained values are summarized in table \ref{table2} where the same parameters obtained by other techniques are also listed. The overall graphene corrugation agrees, within the error bars, with LEED\cite{Hamalainen2013}, AFM\cite{Hamalainen2013}, Surface X-ray Diffraction (SXRD) \cite{Jean2015} and van der Waals-density functional theory (vdW-DFT)\cite{Busse2011} previous studies, but it is lower compared to x-ray standing wave (XSW)\cite{Busse2011}. Moreover, the mean height obtained in this work concurs with all previous structural descriptions listed in \ref{table2}. The concordance of our results with all those techniques reinforces the effectiveness of our simplified description of the problem. For example, a typical periodic model used in this system considers a (10 $\times$ 10) graphene layer on a (9 $\times$ 9) Ir(111) cell. In such a model, we would have to consider 81 nonequivalent emitters for the XPD multiple scattering calculation. However, our result indicates that the chosen emitters already provides an adequate experimental characterization.  
\begin{table}[!htb]
  \begin{center}
    \begin{tabular}{c|c|c}
                                 &$\bar{d}_{C-Ir}$ (\AA) &$\Delta \bar{d}_{C-Ir}$ (\AA)                    \\ \hline \hline
    \textbf{XPD} (this work)                       &      3.40 $\pm$ 0.11         & 0.45 $\pm$ 0.03     \\ 
    \textbf{LEED}\cite{Hamalainen2013} & 3.39 $\pm$ 0.03   & 0.43$\pm$0.09         \\ 
    \textbf{AFM} \cite{Hamalainen2013} &        -           & 0.47$\pm$0.05     \\
    \textbf{SXRD}\cite{Jean2015}       & 3.39 $\pm$ 0.28 & 0.379$\pm$0.044        \\
    \textbf{EXRR} \cite{Jean2015}      & 3.38 $\pm$ 0.04 &       -                 \\
    \textbf{XSW} \cite{Busse2011}      & 3.38 $\pm$ 0.04 & 0.6 $\pm$ 0.1     \\
    \textbf{vdW-DFT} \cite{Busse2011}  & 3.41            & 0.35                      \\
    \end{tabular}
    \caption{Graphene mean height ($\bar{d}_{C-Ir}$) and overall corrugation ($\Delta \bar{d}_{C-Ir}$) calculated by different methodologies.}
    \label{table2}
  \end{center}
\end{table}
\section{Conclusions}
In summary, we have determined the atomic structure of Gr/Ir via XPD. The possibility of discriminating the surface and the bulk iridium photoemission signals allows us to probe the graphene topography using the closest substrate emitters to the carbon atoms. The graphene layer is described by three Gaussian profiles centered on the high-symmetry sites. In the atop region, the carbon atoms are located at the highest distance from the substrate. On the other hand, in the vicinity of the hcp site, the graphene-substrate separation is the shortest. Furthermore,  we obtained the mean graphene height and its overall corrugation, which are in agreement with already reported studies indicating that a proper selection of representative emitters can provide a realistic structural characterization of this system via XPD. This approach might be used to study other long-range periodic systems.  

\begin{acknowledgments}
This work has been supported by  FAPESP  (grant nos. 2016/21402-8,  2017/18574-4, 2007/54829-5, and 2007/08244-5), and CNPq (grant no. 401826/2013-9). We thank  LNLS for the beamtime, and PGM beamline staff for the technical support. The authors are grateful to the Multiuser Central Facilities CEMs-UFABC for the computational support. DCM acknowledges UFABC for the studentship.     
\end{acknowledgments}
\bibliography{subref}

\begin{thebibliography}{59}%
\makeatletter
\providecommand \@ifxundefined [1]{%
 \@ifx{#1\undefined}
}%
\providecommand \@ifnum [1]{%
 \ifnum #1\expandafter \@firstoftwo
 \else \expandafter \@secondoftwo
 \fi
}%
\providecommand \@ifx [1]{%
 \ifx #1\expandafter \@firstoftwo
 \else \expandafter \@secondoftwo
 \fi
}%
\providecommand \natexlab [1]{#1}%
\providecommand \enquote  [1]{``#1''}%
\providecommand \bibnamefont  [1]{#1}%
\providecommand \bibfnamefont [1]{#1}%
\providecommand \citenamefont [1]{#1}%
\providecommand \href@noop [0]{\@secondoftwo}%
\providecommand \href [0]{\begingroup \@sanitize@url \@href}%
\providecommand \@href[1]{\@@startlink{#1}\@@href}%
\providecommand \@@href[1]{\endgroup#1\@@endlink}%
\providecommand \@sanitize@url [0]{\catcode `\\12\catcode `\$12\catcode
  `\&12\catcode `\#12\catcode `\^12\catcode `\_12\catcode `\%12\relax}%
\providecommand \@@startlink[1]{}%
\providecommand \@@endlink[0]{}%
\providecommand \url  [0]{\begingroup\@sanitize@url \@url }%
\providecommand \@url [1]{\endgroup\@href {#1}{\urlprefix }}%
\providecommand \urlprefix  [0]{URL }%
\providecommand \Eprint [0]{\href }%
\providecommand \doibase [0]{http://dx.doi.org/}%
\providecommand \selectlanguage [0]{\@gobble}%
\providecommand \bibinfo  [0]{\@secondoftwo}%
\providecommand \bibfield  [0]{\@secondoftwo}%
\providecommand \translation [1]{[#1]}%
\providecommand \BibitemOpen [0]{}%
\providecommand \bibitemStop [0]{}%
\providecommand \bibitemNoStop [0]{.\EOS\space}%
\providecommand \EOS [0]{\spacefactor3000\relax}%
\providecommand \BibitemShut  [1]{\csname bibitem#1\endcsname}%
\let\auto@bib@innerbib\@empty
\bibitem [{\citenamefont {Novoselov}\ \emph {et~al.}(2004)\citenamefont
  {Novoselov}, \citenamefont {Geim}, \citenamefont {Morozov}, \citenamefont
  {Jiang}, \citenamefont {Zhang}, \citenamefont {Dubonos}, \citenamefont
  {Grigorieva},\ and\ \citenamefont {Firsov}}]{Novoselov2004}%
  \BibitemOpen
  \bibfield  {author} {\bibinfo {author} {\bibfnamefont {K.~S.}\ \bibnamefont
  {Novoselov}}, \bibinfo {author} {\bibfnamefont {A.~K.}\ \bibnamefont {Geim}},
  \bibinfo {author} {\bibfnamefont {S.~V.}\ \bibnamefont {Morozov}}, \bibinfo
  {author} {\bibfnamefont {D.}~\bibnamefont {Jiang}}, \bibinfo {author}
  {\bibfnamefont {Y.}~\bibnamefont {Zhang}}, \bibinfo {author} {\bibfnamefont
  {S.~V.}\ \bibnamefont {Dubonos}}, \bibinfo {author} {\bibfnamefont {I.~V.}\
  \bibnamefont {Grigorieva}}, \ and\ \bibinfo {author} {\bibfnamefont {A.~A.}\
  \bibnamefont {Firsov}},\ }\href {\doibase 10.1126/science.1102896} {\bibfield
   {journal} {\bibinfo  {journal} {Science}\ }\textbf {\bibinfo {volume}
  {306}},\ \bibinfo {pages} {666} (\bibinfo {year} {2004})}\BibitemShut
  {NoStop}%
\bibitem [{\citenamefont {Lee}\ \emph {et~al.}(2019)\citenamefont {Lee},
  \citenamefont {Hiew}, \citenamefont {Lai}, \citenamefont {Lee}, \citenamefont
  {Gan}, \citenamefont {Thangalazhy-Gopakumar},\ and\ \citenamefont
  {Rigby}}]{Lee2019}%
  \BibitemOpen
  \bibfield  {author} {\bibinfo {author} {\bibfnamefont {X.~J.}\ \bibnamefont
  {Lee}}, \bibinfo {author} {\bibfnamefont {B.~Y.~Z.}\ \bibnamefont {Hiew}},
  \bibinfo {author} {\bibfnamefont {K.~C.}\ \bibnamefont {Lai}}, \bibinfo
  {author} {\bibfnamefont {L.~Y.}\ \bibnamefont {Lee}}, \bibinfo {author}
  {\bibfnamefont {S.}~\bibnamefont {Gan}}, \bibinfo {author} {\bibfnamefont
  {S.}~\bibnamefont {Thangalazhy-Gopakumar}}, \ and\ \bibinfo {author}
  {\bibfnamefont {S.}~\bibnamefont {Rigby}},\ }\href {\doibase
  https://doi.org/10.1016/j.jtice.2018.10.028} {\bibfield  {journal} {\bibinfo
  {journal} {Journal of the Taiwan Institute of Chemical Engineers}\ }\textbf
  {\bibinfo {volume} {98}},\ \bibinfo {pages} {163 } (\bibinfo {year}
  {2019})}\BibitemShut {NoStop}%
\bibitem [{\citenamefont {Wintterlin}\ and\ \citenamefont
  {Bocquet}(2009)}]{Wintterlin2009}%
  \BibitemOpen
  \bibfield  {author} {\bibinfo {author} {\bibfnamefont {J.}~\bibnamefont
  {Wintterlin}}\ and\ \bibinfo {author} {\bibfnamefont {M.-L.}\ \bibnamefont
  {Bocquet}},\ }\href {\doibase https://doi.org/10.1016/j.susc.2008.08.037}
  {\bibfield  {journal} {\bibinfo  {journal} {Surface Science}\ }\textbf
  {\bibinfo {volume} {603}},\ \bibinfo {pages} {1841 } (\bibinfo {year}
  {2009})}\BibitemShut {NoStop}%
\bibitem [{\citenamefont {Batzill}(2012)}]{Batzill2012}%
  \BibitemOpen
  \bibfield  {author} {\bibinfo {author} {\bibfnamefont {M.}~\bibnamefont
  {Batzill}},\ }\href {\doibase 10.1016/j.surfrep.2011.12.001} {\bibfield
  {journal} {\bibinfo  {journal} {Surface Science Reports}\ }\textbf {\bibinfo
  {volume} {67}},\ \bibinfo {pages} {83} (\bibinfo {year} {2012})}\BibitemShut
  {NoStop}%
\bibitem [{\citenamefont {Lizzit}\ \emph {et~al.}(2012)\citenamefont {Lizzit},
  \citenamefont {Larciprete}, \citenamefont {Lacovig}, \citenamefont
  {Dalmiglio}, \citenamefont {Orlando}, \citenamefont {Baraldi}, \citenamefont
  {Gammelgaard}, \citenamefont {Barreto}, \citenamefont {Bianchi},
  \citenamefont {Perkins},\ and\ \citenamefont {Hofmann}}]{Lizzit2012}%
  \BibitemOpen
  \bibfield  {author} {\bibinfo {author} {\bibfnamefont {S.}~\bibnamefont
  {Lizzit}}, \bibinfo {author} {\bibfnamefont {R.}~\bibnamefont {Larciprete}},
  \bibinfo {author} {\bibfnamefont {P.}~\bibnamefont {Lacovig}}, \bibinfo
  {author} {\bibfnamefont {M.}~\bibnamefont {Dalmiglio}}, \bibinfo {author}
  {\bibfnamefont {F.}~\bibnamefont {Orlando}}, \bibinfo {author} {\bibfnamefont
  {A.}~\bibnamefont {Baraldi}}, \bibinfo {author} {\bibfnamefont
  {L.}~\bibnamefont {Gammelgaard}}, \bibinfo {author} {\bibfnamefont
  {L.}~\bibnamefont {Barreto}}, \bibinfo {author} {\bibfnamefont
  {M.}~\bibnamefont {Bianchi}}, \bibinfo {author} {\bibfnamefont
  {E.}~\bibnamefont {Perkins}}, \ and\ \bibinfo {author} {\bibfnamefont
  {P.}~\bibnamefont {Hofmann}},\ }\bibfield  {booktitle} {\emph {\bibinfo
  {booktitle} {Nano Letters}},\ }\href {\doibase 10.1021/nl301614j} {\bibfield
  {journal} {\bibinfo  {journal} {Nano Lett.}\ }\textbf {\bibinfo {volume}
  {12}},\ \bibinfo {pages} {4503} (\bibinfo {year} {2012})}\BibitemShut
  {NoStop}%
\bibitem [{\citenamefont {Hattab}\ \emph {et~al.}(2011)\citenamefont {Hattab},
  \citenamefont {N’Diaye}, \citenamefont {Wall}, \citenamefont {Jnawali},
  \citenamefont {Coraux}, \citenamefont {Busse}, \citenamefont {van Gastel},
  \citenamefont {Poelsema}, \citenamefont {Michely}, \citenamefont
  {zu~Heringdorf},\ and\ \citenamefont {von Hoegen}}]{Hattab2011}%
  \BibitemOpen
  \bibfield  {author} {\bibinfo {author} {\bibfnamefont {H.}~\bibnamefont
  {Hattab}}, \bibinfo {author} {\bibfnamefont {A.~T.}\ \bibnamefont
  {N’Diaye}}, \bibinfo {author} {\bibfnamefont {D.}~\bibnamefont {Wall}},
  \bibinfo {author} {\bibfnamefont {G.}~\bibnamefont {Jnawali}}, \bibinfo
  {author} {\bibfnamefont {J.}~\bibnamefont {Coraux}}, \bibinfo {author}
  {\bibfnamefont {C.}~\bibnamefont {Busse}}, \bibinfo {author} {\bibfnamefont
  {R.}~\bibnamefont {van Gastel}}, \bibinfo {author} {\bibfnamefont
  {B.}~\bibnamefont {Poelsema}}, \bibinfo {author} {\bibfnamefont
  {T.}~\bibnamefont {Michely}}, \bibinfo {author} {\bibfnamefont {F.-J.~M.}\
  \bibnamefont {zu~Heringdorf}}, \ and\ \bibinfo {author} {\bibfnamefont
  {M.~H.}\ \bibnamefont {von Hoegen}},\ }\href {\doibase 10.1063/1.3548546}
  {\bibfield  {journal} {\bibinfo  {journal} {Applied Physics Letters}\
  }\textbf {\bibinfo {volume} {98}},\ \bibinfo {pages} {141903} (\bibinfo
  {year} {2011})}\BibitemShut {NoStop}%
\bibitem [{\citenamefont {Pletikosi\ifmmode~\acute{c}\else \'{c}\fi{}}\ \emph
  {et~al.}(2009)\citenamefont {Pletikosi\ifmmode~\acute{c}\else \'{c}\fi{}},
  \citenamefont {Kralj}, \citenamefont {Pervan}, \citenamefont {Brako},
  \citenamefont {Coraux}, \citenamefont {N'Diaye}, \citenamefont {Busse},\ and\
  \citenamefont {Michely}}]{Pletikosic2009}%
  \BibitemOpen
  \bibfield  {author} {\bibinfo {author} {\bibfnamefont {I.}~\bibnamefont
  {Pletikosi\ifmmode~\acute{c}\else \'{c}\fi{}}}, \bibinfo {author}
  {\bibfnamefont {M.}~\bibnamefont {Kralj}}, \bibinfo {author} {\bibfnamefont
  {P.}~\bibnamefont {Pervan}}, \bibinfo {author} {\bibfnamefont
  {R.}~\bibnamefont {Brako}}, \bibinfo {author} {\bibfnamefont
  {J.}~\bibnamefont {Coraux}}, \bibinfo {author} {\bibfnamefont {A.~T.}\
  \bibnamefont {N'Diaye}}, \bibinfo {author} {\bibfnamefont {C.}~\bibnamefont
  {Busse}}, \ and\ \bibinfo {author} {\bibfnamefont {T.}~\bibnamefont
  {Michely}},\ }\href {\doibase 10.1103/PhysRevLett.102.056808} {\bibfield
  {journal} {\bibinfo  {journal} {Phys. Rev. Lett.}\ }\textbf {\bibinfo
  {volume} {102}},\ \bibinfo {pages} {056808} (\bibinfo {year}
  {2009})}\BibitemShut {NoStop}%
\bibitem [{\citenamefont {Kralj}\ \emph {et~al.}(2011)\citenamefont {Kralj},
  \citenamefont {Pletikosi\ifmmode~\acute{c}\else \'{c}\fi{}}, \citenamefont
  {Petrovi\ifmmode~\acute{c}\else \'{c}\fi{}}, \citenamefont {Pervan},
  \citenamefont {Milun}, \citenamefont {N'Diaye}, \citenamefont {Busse},
  \citenamefont {Michely}, \citenamefont {Fujii},\ and\ \citenamefont
  {Vobornik}}]{kralj2011}%
  \BibitemOpen
  \bibfield  {author} {\bibinfo {author} {\bibfnamefont {M.}~\bibnamefont
  {Kralj}}, \bibinfo {author} {\bibfnamefont {I.}~\bibnamefont
  {Pletikosi\ifmmode~\acute{c}\else \'{c}\fi{}}}, \bibinfo {author}
  {\bibfnamefont {M.}~\bibnamefont {Petrovi\ifmmode~\acute{c}\else
  \'{c}\fi{}}}, \bibinfo {author} {\bibfnamefont {P.}~\bibnamefont {Pervan}},
  \bibinfo {author} {\bibfnamefont {M.}~\bibnamefont {Milun}}, \bibinfo
  {author} {\bibfnamefont {A.~T.}\ \bibnamefont {N'Diaye}}, \bibinfo {author}
  {\bibfnamefont {C.}~\bibnamefont {Busse}}, \bibinfo {author} {\bibfnamefont
  {T.}~\bibnamefont {Michely}}, \bibinfo {author} {\bibfnamefont
  {J.}~\bibnamefont {Fujii}}, \ and\ \bibinfo {author} {\bibfnamefont
  {I.}~\bibnamefont {Vobornik}},\ }\href {\doibase 10.1103/PhysRevB.84.075427}
  {\bibfield  {journal} {\bibinfo  {journal} {Phys. Rev. B}\ }\textbf {\bibinfo
  {volume} {84}},\ \bibinfo {pages} {075427} (\bibinfo {year}
  {2011})}\BibitemShut {NoStop}%
\bibitem [{\citenamefont {Larciprete}\ \emph {et~al.}(2012)\citenamefont
  {Larciprete}, \citenamefont {Ulstrup}, \citenamefont {Lacovig}, \citenamefont
  {Dalmiglio}, \citenamefont {Bianchi}, \citenamefont {Mazzola}, \citenamefont
  {Horne{\ae}r}, \citenamefont {Orlando}, \citenamefont {Baraldi},
  \citenamefont {Hofmann},\ and\ \citenamefont {Lizzit}}]{Larciprete2012}%
  \BibitemOpen
  \bibfield  {author} {\bibinfo {author} {\bibfnamefont {R.}~\bibnamefont
  {Larciprete}}, \bibinfo {author} {\bibfnamefont {S.}~\bibnamefont {Ulstrup}},
  \bibinfo {author} {\bibfnamefont {P.}~\bibnamefont {Lacovig}}, \bibinfo
  {author} {\bibfnamefont {M.}~\bibnamefont {Dalmiglio}}, \bibinfo {author}
  {\bibfnamefont {M.}~\bibnamefont {Bianchi}}, \bibinfo {author} {\bibfnamefont
  {F.}~\bibnamefont {Mazzola}}, \bibinfo {author} {\bibfnamefont
  {L.}~\bibnamefont {Horne{\ae}r}}, \bibinfo {author} {\bibfnamefont
  {F.}~\bibnamefont {Orlando}}, \bibinfo {author} {\bibfnamefont
  {A.}~\bibnamefont {Baraldi}}, \bibinfo {author} {\bibfnamefont
  {P.}~\bibnamefont {Hofmann}}, \ and\ \bibinfo {author} {\bibfnamefont
  {S.}~\bibnamefont {Lizzit}},\ }\href {\doibase 10.1021/nn302729j} {\bibfield
  {journal} {\bibinfo  {journal} {ACS Nano}\ }\textbf {\bibinfo {volume} {6}},\
  \bibinfo {pages} {9551} (\bibinfo {year} {2012})}\BibitemShut {NoStop}%
\bibitem [{\citenamefont {Ulstrup}\ \emph {et~al.}(2014)\citenamefont
  {Ulstrup}, \citenamefont {Andersen}, \citenamefont {Bianchi}, \citenamefont
  {Barreto}, \citenamefont {Hammer}, \citenamefont {Hornekær},\ and\
  \citenamefont {Hofmann}}]{Ulstrup2014}%
  \BibitemOpen
  \bibfield  {author} {\bibinfo {author} {\bibfnamefont {S.}~\bibnamefont
  {Ulstrup}}, \bibinfo {author} {\bibfnamefont {M.}~\bibnamefont {Andersen}},
  \bibinfo {author} {\bibfnamefont {M.}~\bibnamefont {Bianchi}}, \bibinfo
  {author} {\bibfnamefont {L.}~\bibnamefont {Barreto}}, \bibinfo {author}
  {\bibfnamefont {B.}~\bibnamefont {Hammer}}, \bibinfo {author} {\bibfnamefont
  {L.}~\bibnamefont {Hornekær}}, \ and\ \bibinfo {author} {\bibfnamefont
  {P.}~\bibnamefont {Hofmann}},\ }\href
  {http://stacks.iop.org/2053-1583/1/i=2/a=025002} {\bibfield  {journal}
  {\bibinfo  {journal} {2D Materials}\ }\textbf {\bibinfo {volume} {1}},\
  \bibinfo {pages} {025002} (\bibinfo {year} {2014})}\BibitemShut {NoStop}%
\bibitem [{\citenamefont {Pervan}\ and\ \citenamefont
  {Lazi\ifmmode~\acute{c}\else \'{c}\fi{}}(2017)}]{Pervan2017}%
  \BibitemOpen
  \bibfield  {author} {\bibinfo {author} {\bibfnamefont {P.}~\bibnamefont
  {Pervan}}\ and\ \bibinfo {author} {\bibfnamefont {P.}~\bibnamefont
  {Lazi\ifmmode~\acute{c}\else \'{c}\fi{}}},\ }\href {\doibase
  10.1103/PhysRevMaterials.1.044202} {\bibfield  {journal} {\bibinfo  {journal}
  {Phys. Rev. Materials}\ }\textbf {\bibinfo {volume} {1}},\ \bibinfo {pages}
  {044202} (\bibinfo {year} {2017})}\BibitemShut {NoStop}%
\bibitem [{\citenamefont {Balog}\ \emph {et~al.}(2019)\citenamefont {Balog},
  \citenamefont {Cassidy}, \citenamefont {J{\o}rgensen}, \citenamefont {Kyhl},
  \citenamefont {Andersen}, \citenamefont {{\v{C}}abo}, \citenamefont {Ravani},
  \citenamefont {Bignardi}, \citenamefont {Lacovig}, \citenamefont {Lizzit},\
  and\ \citenamefont {Hornek{\ae}r}}]{Balog2019}%
  \BibitemOpen
  \bibfield  {author} {\bibinfo {author} {\bibfnamefont {R.}~\bibnamefont
  {Balog}}, \bibinfo {author} {\bibfnamefont {A.}~\bibnamefont {Cassidy}},
  \bibinfo {author} {\bibfnamefont {J.}~\bibnamefont {J{\o}rgensen}}, \bibinfo
  {author} {\bibfnamefont {L.}~\bibnamefont {Kyhl}}, \bibinfo {author}
  {\bibfnamefont {M.}~\bibnamefont {Andersen}}, \bibinfo {author}
  {\bibfnamefont {A.~G.}\ \bibnamefont {{\v{C}}abo}}, \bibinfo {author}
  {\bibfnamefont {F.}~\bibnamefont {Ravani}}, \bibinfo {author} {\bibfnamefont
  {L.}~\bibnamefont {Bignardi}}, \bibinfo {author} {\bibfnamefont
  {P.}~\bibnamefont {Lacovig}}, \bibinfo {author} {\bibfnamefont
  {S.}~\bibnamefont {Lizzit}}, \ and\ \bibinfo {author} {\bibfnamefont
  {L.}~\bibnamefont {Hornek{\ae}r}},\ }\href {\doibase
  10.1088/1361-648x/aaf76b} {\bibfield  {journal} {\bibinfo  {journal} {Journal
  of Physics: Condensed Matter}\ }\textbf {\bibinfo {volume} {31}},\ \bibinfo
  {pages} {085001} (\bibinfo {year} {2019})}\BibitemShut {NoStop}%
\bibitem [{\citenamefont {Dedkov}\ and\ \citenamefont
  {Voloshina}(2015)}]{Dedkov2015}%
  \BibitemOpen
  \bibfield  {author} {\bibinfo {author} {\bibfnamefont {Y.}~\bibnamefont
  {Dedkov}}\ and\ \bibinfo {author} {\bibfnamefont {E.}~\bibnamefont
  {Voloshina}},\ }\href {http://stacks.iop.org/0953-8984/27/i=30/a=303002}
  {\bibfield  {journal} {\bibinfo  {journal} {Journal of Physics: Condensed
  Matter}\ }\textbf {\bibinfo {volume} {27}},\ \bibinfo {pages} {303002}
  (\bibinfo {year} {2015})}\BibitemShut {NoStop}%
\bibitem [{\citenamefont {Voloshina}\ \emph {et~al.}(2013)\citenamefont
  {Voloshina}, \citenamefont {Fertitta}, \citenamefont {Garhofer},
  \citenamefont {Mittendorfer}, \citenamefont {Fonin}, \citenamefont
  {Thissen},\ and\ \citenamefont {Dedkov}}]{Voloshina2013}%
  \BibitemOpen
  \bibfield  {author} {\bibinfo {author} {\bibfnamefont {E.~N.}\ \bibnamefont
  {Voloshina}}, \bibinfo {author} {\bibfnamefont {E.}~\bibnamefont {Fertitta}},
  \bibinfo {author} {\bibfnamefont {A.}~\bibnamefont {Garhofer}}, \bibinfo
  {author} {\bibfnamefont {F.}~\bibnamefont {Mittendorfer}}, \bibinfo {author}
  {\bibfnamefont {M.}~\bibnamefont {Fonin}}, \bibinfo {author} {\bibfnamefont
  {A.}~\bibnamefont {Thissen}}, \ and\ \bibinfo {author} {\bibfnamefont
  {Y.~S.}\ \bibnamefont {Dedkov}},\ }\href {https://doi.org/10.1038/srep01072}
  {\bibfield  {journal} {\bibinfo  {journal} {Scientific Reports}\ }\textbf
  {\bibinfo {volume} {3}},\ \bibinfo {pages} {1072} (\bibinfo {year}
  {2013})}\BibitemShut {NoStop}%
\bibitem [{\citenamefont {N'Diaye}\ \emph {et~al.}(2006)\citenamefont
  {N'Diaye}, \citenamefont {Bleikamp}, \citenamefont {Feibelman},\ and\
  \citenamefont {Michely}}]{Ndiaye2006}%
  \BibitemOpen
  \bibfield  {author} {\bibinfo {author} {\bibfnamefont {A.~T.}\ \bibnamefont
  {N'Diaye}}, \bibinfo {author} {\bibfnamefont {S.}~\bibnamefont {Bleikamp}},
  \bibinfo {author} {\bibfnamefont {P.~J.}\ \bibnamefont {Feibelman}}, \ and\
  \bibinfo {author} {\bibfnamefont {T.}~\bibnamefont {Michely}},\ }\href
  {\doibase 10.1103/PhysRevLett.97.215501} {\bibfield  {journal} {\bibinfo
  {journal} {Phys. Rev. Lett.}\ }\textbf {\bibinfo {volume} {97}},\ \bibinfo
  {pages} {215501} (\bibinfo {year} {2006})}\BibitemShut {NoStop}%
\bibitem [{\citenamefont {Baltic}\ \emph {et~al.}(2016)\citenamefont {Baltic},
  \citenamefont {Pivetta}, \citenamefont {Donati}, \citenamefont {W\"ackerlin},
  \citenamefont {Singha}, \citenamefont {Dreiser}, \citenamefont {Rusponi},\
  and\ \citenamefont {Brune}}]{Baltic2016}%
  \BibitemOpen
  \bibfield  {author} {\bibinfo {author} {\bibfnamefont {R.}~\bibnamefont
  {Baltic}}, \bibinfo {author} {\bibfnamefont {M.}~\bibnamefont {Pivetta}},
  \bibinfo {author} {\bibfnamefont {F.}~\bibnamefont {Donati}}, \bibinfo
  {author} {\bibfnamefont {C.}~\bibnamefont {W\"ackerlin}}, \bibinfo {author}
  {\bibfnamefont {A.}~\bibnamefont {Singha}}, \bibinfo {author} {\bibfnamefont
  {J.}~\bibnamefont {Dreiser}}, \bibinfo {author} {\bibfnamefont
  {S.}~\bibnamefont {Rusponi}}, \ and\ \bibinfo {author} {\bibfnamefont
  {H.}~\bibnamefont {Brune}},\ }\href {\doibase 10.1021/acs.nanolett.6b03543}
  {\bibfield  {journal} {\bibinfo  {journal} {Nano Letters}\ }\textbf {\bibinfo
  {volume} {16}},\ \bibinfo {pages} {7610} (\bibinfo {year}
  {2016})}\BibitemShut {NoStop}%
\bibitem [{\citenamefont {Petrovi\ifmmode~\acute{c}\else \'{c}\fi{}}\ \emph
  {et~al.}(2017)\citenamefont {Petrovi\ifmmode~\acute{c}\else \'{c}\fi{}},
  \citenamefont {Lazi\ifmmode~\acute{c}\else \'{c}\fi{}}, \citenamefont
  {Runte}, \citenamefont {Michely}, \citenamefont {Busse},\ and\ \citenamefont
  {Kralj}}]{Petrovic2017}%
  \BibitemOpen
  \bibfield  {author} {\bibinfo {author} {\bibfnamefont {M.}~\bibnamefont
  {Petrovi\ifmmode~\acute{c}\else \'{c}\fi{}}}, \bibinfo {author}
  {\bibfnamefont {P.}~\bibnamefont {Lazi\ifmmode~\acute{c}\else \'{c}\fi{}}},
  \bibinfo {author} {\bibfnamefont {S.}~\bibnamefont {Runte}}, \bibinfo
  {author} {\bibfnamefont {T.}~\bibnamefont {Michely}}, \bibinfo {author}
  {\bibfnamefont {C.}~\bibnamefont {Busse}}, \ and\ \bibinfo {author}
  {\bibfnamefont {M.}~\bibnamefont {Kralj}},\ }\href {\doibase
  10.1103/PhysRevB.96.085428} {\bibfield  {journal} {\bibinfo  {journal} {Phys.
  Rev. B}\ }\textbf {\bibinfo {volume} {96}},\ \bibinfo {pages} {085428}
  (\bibinfo {year} {2017})}\BibitemShut {NoStop}%
\bibitem [{\citenamefont {Pivetta}\ \emph {et~al.}(2018)\citenamefont
  {Pivetta}, \citenamefont {Rusponi},\ and\ \citenamefont
  {Brune}}]{Pivetta2018}%
  \BibitemOpen
  \bibfield  {author} {\bibinfo {author} {\bibfnamefont {M.}~\bibnamefont
  {Pivetta}}, \bibinfo {author} {\bibfnamefont {S.}~\bibnamefont {Rusponi}}, \
  and\ \bibinfo {author} {\bibfnamefont {H.}~\bibnamefont {Brune}},\ }\href
  {\doibase 10.1103/PhysRevB.98.115417} {\bibfield  {journal} {\bibinfo
  {journal} {Phys. Rev. B}\ }\textbf {\bibinfo {volume} {98}},\ \bibinfo
  {pages} {115417} (\bibinfo {year} {2018})}\BibitemShut {NoStop}%
\bibitem [{\citenamefont {S{\o}rensen}\ \emph {et~al.}(2014)\citenamefont
  {S{\o}rensen}, \citenamefont {F\"{u}chtbauer}, \citenamefont {Tuxen},
  \citenamefont {Walton},\ and\ \citenamefont {Lauritsen}}]{Sorensen2014}%
  \BibitemOpen
  \bibfield  {author} {\bibinfo {author} {\bibfnamefont {S.~G.}\ \bibnamefont
  {S{\o}rensen}}, \bibinfo {author} {\bibfnamefont {H.~G.}\ \bibnamefont
  {F\"{u}chtbauer}}, \bibinfo {author} {\bibfnamefont {A.~K.}\ \bibnamefont
  {Tuxen}}, \bibinfo {author} {\bibfnamefont {A.~S.}\ \bibnamefont {Walton}}, \
  and\ \bibinfo {author} {\bibfnamefont {J.~V.}\ \bibnamefont {Lauritsen}},\
  }\href {\doibase 10.1021/nn502812n} {\bibfield  {journal} {\bibinfo
  {journal} {ACS Nano}\ }\textbf {\bibinfo {volume} {8}},\ \bibinfo {pages}
  {6788} (\bibinfo {year} {2014})}\BibitemShut {NoStop}%
\bibitem [{\citenamefont {Gr{\o}nborg}\ \emph {et~al.}(2015)\citenamefont
  {Gr{\o}nborg}, \citenamefont {Ulstrup}, \citenamefont {Bianchi},
  \citenamefont {Dendzik}, \citenamefont {Sanders}, \citenamefont {Lauritsen},
  \citenamefont {Hofmann},\ and\ \citenamefont {Miwa}}]{Gronborg2015}%
  \BibitemOpen
  \bibfield  {author} {\bibinfo {author} {\bibfnamefont {S.~S.}\ \bibnamefont
  {Gr{\o}nborg}}, \bibinfo {author} {\bibfnamefont {S.}~\bibnamefont
  {Ulstrup}}, \bibinfo {author} {\bibfnamefont {M.}~\bibnamefont {Bianchi}},
  \bibinfo {author} {\bibfnamefont {M.}~\bibnamefont {Dendzik}}, \bibinfo
  {author} {\bibfnamefont {C.~E.}\ \bibnamefont {Sanders}}, \bibinfo {author}
  {\bibfnamefont {J.~V.}\ \bibnamefont {Lauritsen}}, \bibinfo {author}
  {\bibfnamefont {P.}~\bibnamefont {Hofmann}}, \ and\ \bibinfo {author}
  {\bibfnamefont {J.~A.}\ \bibnamefont {Miwa}},\ }\href {\doibase
  10.1021/acs.langmuir.5b02533} {\bibfield  {journal} {\bibinfo  {journal}
  {Langmuir}\ }\textbf {\bibinfo {volume} {31}},\ \bibinfo {pages} {9700}
  (\bibinfo {year} {2015})}\BibitemShut {NoStop}%
\bibitem [{\citenamefont {Dendzik}\ \emph {et~al.}(2015)\citenamefont
  {Dendzik}, \citenamefont {Michiardi}, \citenamefont {Sanders}, \citenamefont
  {Bianchi}, \citenamefont {Miwa}, \citenamefont {Gr\o{}nborg}, \citenamefont
  {Lauritsen}, \citenamefont {Bruix}, \citenamefont {Hammer},\ and\
  \citenamefont {Hofmann}}]{Dendzik2015}%
  \BibitemOpen
  \bibfield  {author} {\bibinfo {author} {\bibfnamefont {M.}~\bibnamefont
  {Dendzik}}, \bibinfo {author} {\bibfnamefont {M.}~\bibnamefont {Michiardi}},
  \bibinfo {author} {\bibfnamefont {C.}~\bibnamefont {Sanders}}, \bibinfo
  {author} {\bibfnamefont {M.}~\bibnamefont {Bianchi}}, \bibinfo {author}
  {\bibfnamefont {J.~A.}\ \bibnamefont {Miwa}}, \bibinfo {author}
  {\bibfnamefont {S.~S.}\ \bibnamefont {Gr\o{}nborg}}, \bibinfo {author}
  {\bibfnamefont {J.~V.}\ \bibnamefont {Lauritsen}}, \bibinfo {author}
  {\bibfnamefont {A.}~\bibnamefont {Bruix}}, \bibinfo {author} {\bibfnamefont
  {B.}~\bibnamefont {Hammer}}, \ and\ \bibinfo {author} {\bibfnamefont
  {P.}~\bibnamefont {Hofmann}},\ }\href {\doibase 10.1103/PhysRevB.92.245442}
  {\bibfield  {journal} {\bibinfo  {journal} {Phys. Rev. B}\ }\textbf {\bibinfo
  {volume} {92}},\ \bibinfo {pages} {245442} (\bibinfo {year}
  {2015})}\BibitemShut {NoStop}%
\bibitem [{\citenamefont {Auw\"{a}rter}(2019)}]{Auwarter2019}%
  \BibitemOpen
  \bibfield  {author} {\bibinfo {author} {\bibfnamefont {W.}~\bibnamefont
  {Auw\"{a}rter}},\ }\href {\doibase
  https://doi.org/10.1016/j.surfrep.2018.10.001} {\bibfield  {journal}
  {\bibinfo  {journal} {Surface Science Reports}\ }\textbf {\bibinfo {volume}
  {74}},\ \bibinfo {pages} {1 } (\bibinfo {year} {2019})}\BibitemShut {NoStop}%
\bibitem [{\citenamefont {Cheng}\ \emph {et~al.}(2019)\citenamefont {Cheng},
  \citenamefont {Huang}, \citenamefont {Hong}, \citenamefont {Zhao},\ and\
  \citenamefont {Liu}}]{Cheng2019}%
  \BibitemOpen
  \bibfield  {author} {\bibinfo {author} {\bibfnamefont {Y.}~\bibnamefont
  {Cheng}}, \bibinfo {author} {\bibfnamefont {C.}~\bibnamefont {Huang}},
  \bibinfo {author} {\bibfnamefont {H.}~\bibnamefont {Hong}}, \bibinfo {author}
  {\bibfnamefont {Z.}~\bibnamefont {Zhao}}, \ and\ \bibinfo {author}
  {\bibfnamefont {K.}~\bibnamefont {Liu}},\ }\href {\doibase
  10.1088/1674-1056/ab3e46} {\bibfield  {journal} {\bibinfo  {journal} {Chinese
  Physics B}\ }\textbf {\bibinfo {volume} {28}},\ \bibinfo {pages} {107304}
  (\bibinfo {year} {2019})}\BibitemShut {NoStop}%
\bibitem [{\citenamefont {Cao}\ \emph {et~al.}(2018)\citenamefont {Cao},
  \citenamefont {Fatemi}, \citenamefont {Demir}, \citenamefont {Fang},
  \citenamefont {Tomarken}, \citenamefont {Luo}, \citenamefont
  {Sanchez-Yamagishi}, \citenamefont {Watanabe}, \citenamefont {Taniguchi},
  \citenamefont {Kaxiras}, \citenamefont {Ashoori},\ and\ \citenamefont
  {Jarillo-Herrero}}]{Cao2018}%
  \BibitemOpen
  \bibfield  {author} {\bibinfo {author} {\bibfnamefont {Y.}~\bibnamefont
  {Cao}}, \bibinfo {author} {\bibfnamefont {V.}~\bibnamefont {Fatemi}},
  \bibinfo {author} {\bibfnamefont {A.}~\bibnamefont {Demir}}, \bibinfo
  {author} {\bibfnamefont {S.}~\bibnamefont {Fang}}, \bibinfo {author}
  {\bibfnamefont {S.~L.}\ \bibnamefont {Tomarken}}, \bibinfo {author}
  {\bibfnamefont {J.~Y.}\ \bibnamefont {Luo}}, \bibinfo {author} {\bibfnamefont
  {J.~D.}\ \bibnamefont {Sanchez-Yamagishi}}, \bibinfo {author} {\bibfnamefont
  {K.}~\bibnamefont {Watanabe}}, \bibinfo {author} {\bibfnamefont
  {T.}~\bibnamefont {Taniguchi}}, \bibinfo {author} {\bibfnamefont
  {E.}~\bibnamefont {Kaxiras}}, \bibinfo {author} {\bibfnamefont {R.~C.}\
  \bibnamefont {Ashoori}}, \ and\ \bibinfo {author} {\bibfnamefont
  {P.}~\bibnamefont {Jarillo-Herrero}},\ }\href
  {https://doi.org/10.1038/nature26154} {\bibfield  {journal} {\bibinfo
  {journal} {Nature}\ }\textbf {\bibinfo {volume} {556}},\ \bibinfo {pages}
  {80} (\bibinfo {year} {2018})}\BibitemShut {NoStop}%
\bibitem [{\citenamefont {Zhao}\ \emph {et~al.}(2020)\citenamefont {Zhao},
  \citenamefont {Yang}, \citenamefont {Zhang},\ and\ \citenamefont
  {Wei}}]{Zhao2020}%
  \BibitemOpen
  \bibfield  {author} {\bibinfo {author} {\bibfnamefont {X.-J.}\ \bibnamefont
  {Zhao}}, \bibinfo {author} {\bibfnamefont {Y.}~\bibnamefont {Yang}}, \bibinfo
  {author} {\bibfnamefont {D.-B.}\ \bibnamefont {Zhang}}, \ and\ \bibinfo
  {author} {\bibfnamefont {S.-H.}\ \bibnamefont {Wei}},\ }\href {\doibase
  10.1103/PhysRevLett.124.086401} {\bibfield  {journal} {\bibinfo  {journal}
  {Phys. Rev. Lett.}\ }\textbf {\bibinfo {volume} {124}},\ \bibinfo {pages}
  {086401} (\bibinfo {year} {2020})}\BibitemShut {NoStop}%
\bibitem [{\citenamefont {H\"{a}m\"{a}l\"{a}inen}\ \emph
  {et~al.}(2013)\citenamefont {H\"{a}m\"{a}l\"{a}inen}, \citenamefont
  {Boneschanscher}, \citenamefont {Jacobse}, \citenamefont {Swart},
  \citenamefont {Pussi}, \citenamefont {Moritz}, \citenamefont {Lahtinen},
  \citenamefont {Liljeroth},\ and\ \citenamefont {Sainio}}]{Hamalainen2013}%
  \BibitemOpen
  \bibfield  {author} {\bibinfo {author} {\bibfnamefont {S.~K.}\ \bibnamefont
  {H\"{a}m\"{a}l\"{a}inen}}, \bibinfo {author} {\bibfnamefont {M.~P.}\
  \bibnamefont {Boneschanscher}}, \bibinfo {author} {\bibfnamefont {P.~H.}\
  \bibnamefont {Jacobse}}, \bibinfo {author} {\bibfnamefont {I.}~\bibnamefont
  {Swart}}, \bibinfo {author} {\bibfnamefont {K.}~\bibnamefont {Pussi}},
  \bibinfo {author} {\bibfnamefont {W.}~\bibnamefont {Moritz}}, \bibinfo
  {author} {\bibfnamefont {J.}~\bibnamefont {Lahtinen}}, \bibinfo {author}
  {\bibfnamefont {P.}~\bibnamefont {Liljeroth}}, \ and\ \bibinfo {author}
  {\bibfnamefont {J.}~\bibnamefont {Sainio}},\ }\href {\doibase
  10.1103/PhysRevB.88.201406} {\bibfield  {journal} {\bibinfo  {journal} {Phys.
  Rev. B}\ }\textbf {\bibinfo {volume} {88}},\ \bibinfo {pages} {201406}
  (\bibinfo {year} {2013})}\BibitemShut {NoStop}%
\bibitem [{\citenamefont {N{\textquotesingle}Diaye}\ \emph
  {et~al.}(2008)\citenamefont {N{\textquotesingle}Diaye}, \citenamefont
  {Coraux}, \citenamefont {Plasa}, \citenamefont {Busse},\ and\ \citenamefont
  {Michely}}]{Diaye2008}%
  \BibitemOpen
  \bibfield  {author} {\bibinfo {author} {\bibfnamefont {A.~T.}\ \bibnamefont
  {N{\textquotesingle}Diaye}}, \bibinfo {author} {\bibfnamefont
  {J.}~\bibnamefont {Coraux}}, \bibinfo {author} {\bibfnamefont {T.~N.}\
  \bibnamefont {Plasa}}, \bibinfo {author} {\bibfnamefont {C.}~\bibnamefont
  {Busse}}, \ and\ \bibinfo {author} {\bibfnamefont {T.}~\bibnamefont
  {Michely}},\ }\href {\doibase 10.1088/1367-2630/10/4/043033} {\bibfield
  {journal} {\bibinfo  {journal} {New Journal of Physics}\ }\textbf {\bibinfo
  {volume} {10}},\ \bibinfo {pages} {043033} (\bibinfo {year}
  {2008})}\BibitemShut {NoStop}%
\bibitem [{\citenamefont {de~Campos~Ferreira}\ \emph
  {et~al.}(2018)\citenamefont {de~Campos~Ferreira}, \citenamefont {de~Lima},
  \citenamefont {Barreto}, \citenamefont {Silva}, \citenamefont {Landers},\
  and\ \citenamefont {de~Siervo}}]{Ferreira2018}%
  \BibitemOpen
  \bibfield  {author} {\bibinfo {author} {\bibfnamefont {R.~C.}\ \bibnamefont
  {de~Campos~Ferreira}}, \bibinfo {author} {\bibfnamefont {L.~H.}\ \bibnamefont
  {de~Lima}}, \bibinfo {author} {\bibfnamefont {L.}~\bibnamefont {Barreto}},
  \bibinfo {author} {\bibfnamefont {C.~C.}\ \bibnamefont {Silva}}, \bibinfo
  {author} {\bibfnamefont {R.}~\bibnamefont {Landers}}, \ and\ \bibinfo
  {author} {\bibfnamefont {A.}~\bibnamefont {de~Siervo}},\ }\href {\doibase
  10.1021/acs.chemmater.8b03186} {\bibfield  {journal} {\bibinfo  {journal}
  {Chemistry of Materials}\ }\textbf {\bibinfo {volume} {30}},\ \bibinfo
  {pages} {7201} (\bibinfo {year} {2018})}\BibitemShut {NoStop}%
\bibitem [{\citenamefont {Busse}\ \emph {et~al.}(2011)\citenamefont {Busse},
  \citenamefont {Lazi\ifmmode~\acute{c}\else \'{c}\fi{}}, \citenamefont
  {Djemour}, \citenamefont {Coraux}, \citenamefont {Gerber}, \citenamefont
  {Atodiresei}, \citenamefont {Caciuc}, \citenamefont {Brako}, \citenamefont
  {N'Diaye}, \citenamefont {Bl\"ugel}, \citenamefont {Zegenhagen},\ and\
  \citenamefont {Michely}}]{Busse2011}%
  \BibitemOpen
  \bibfield  {author} {\bibinfo {author} {\bibfnamefont {C.}~\bibnamefont
  {Busse}}, \bibinfo {author} {\bibfnamefont {P.}~\bibnamefont
  {Lazi\ifmmode~\acute{c}\else \'{c}\fi{}}}, \bibinfo {author} {\bibfnamefont
  {R.}~\bibnamefont {Djemour}}, \bibinfo {author} {\bibfnamefont
  {J.}~\bibnamefont {Coraux}}, \bibinfo {author} {\bibfnamefont
  {T.}~\bibnamefont {Gerber}}, \bibinfo {author} {\bibfnamefont
  {N.}~\bibnamefont {Atodiresei}}, \bibinfo {author} {\bibfnamefont
  {V.}~\bibnamefont {Caciuc}}, \bibinfo {author} {\bibfnamefont
  {R.}~\bibnamefont {Brako}}, \bibinfo {author} {\bibfnamefont {A.~T.}\
  \bibnamefont {N'Diaye}}, \bibinfo {author} {\bibfnamefont {S.}~\bibnamefont
  {Bl\"ugel}}, \bibinfo {author} {\bibfnamefont {J.}~\bibnamefont
  {Zegenhagen}}, \ and\ \bibinfo {author} {\bibfnamefont {T.}~\bibnamefont
  {Michely}},\ }\href {\doibase 10.1103/PhysRevLett.107.036101} {\bibfield
  {journal} {\bibinfo  {journal} {Phys. Rev. Lett.}\ }\textbf {\bibinfo
  {volume} {107}},\ \bibinfo {pages} {036101} (\bibinfo {year}
  {2011})}\BibitemShut {NoStop}%
\bibitem [{\citenamefont {Jean}\ \emph {et~al.}(2015)\citenamefont {Jean},
  \citenamefont {Zhou}, \citenamefont {Blanc}, \citenamefont {Felici},
  \citenamefont {Coraux},\ and\ \citenamefont {Renaud}}]{Jean2015}%
  \BibitemOpen
  \bibfield  {author} {\bibinfo {author} {\bibfnamefont {F.}~\bibnamefont
  {Jean}}, \bibinfo {author} {\bibfnamefont {T.}~\bibnamefont {Zhou}}, \bibinfo
  {author} {\bibfnamefont {N.}~\bibnamefont {Blanc}}, \bibinfo {author}
  {\bibfnamefont {R.}~\bibnamefont {Felici}}, \bibinfo {author} {\bibfnamefont
  {J.}~\bibnamefont {Coraux}}, \ and\ \bibinfo {author} {\bibfnamefont
  {G.}~\bibnamefont {Renaud}},\ }\href {\doibase 10.1103/PhysRevB.91.245424}
  {\bibfield  {journal} {\bibinfo  {journal} {Phys. Rev. B}\ }\textbf {\bibinfo
  {volume} {91}},\ \bibinfo {pages} {245424} (\bibinfo {year}
  {2015})}\BibitemShut {NoStop}%
\bibitem [{\citenamefont {Rogge}\ \emph {et~al.}(2015)\citenamefont {Rogge},
  \citenamefont {Th\"{u}rmer}, \citenamefont {Foster}, \citenamefont {McCarty},
  \citenamefont {Dubon},\ and\ \citenamefont {Bartelt}}]{Rogge2015}%
  \BibitemOpen
  \bibfield  {author} {\bibinfo {author} {\bibfnamefont {P.~C.}\ \bibnamefont
  {Rogge}}, \bibinfo {author} {\bibfnamefont {K.}~\bibnamefont {Th\"{u}rmer}},
  \bibinfo {author} {\bibfnamefont {M.~E.}\ \bibnamefont {Foster}}, \bibinfo
  {author} {\bibfnamefont {K.~F.}\ \bibnamefont {McCarty}}, \bibinfo {author}
  {\bibfnamefont {O.~D.}\ \bibnamefont {Dubon}}, \ and\ \bibinfo {author}
  {\bibfnamefont {N.~C.}\ \bibnamefont {Bartelt}},\ }\href
  {https://doi.org/10.1038/ncomms7880} {\bibfield  {journal} {\bibinfo
  {journal} {Nature Communications}\ }\textbf {\bibinfo {volume} {6}},\
  \bibinfo {pages} {6880} (\bibinfo {year} {2015})}\BibitemShut {NoStop}%
\bibitem [{\citenamefont {Moritz}\ \emph {et~al.}(2010)\citenamefont {Moritz},
  \citenamefont {Wang}, \citenamefont {Bocquet}, \citenamefont {Brugger},
  \citenamefont {Greber}, \citenamefont {Wintterlin},\ and\ \citenamefont
  {G\"unther}}]{Moritz2010}%
  \BibitemOpen
  \bibfield  {author} {\bibinfo {author} {\bibfnamefont {W.}~\bibnamefont
  {Moritz}}, \bibinfo {author} {\bibfnamefont {B.}~\bibnamefont {Wang}},
  \bibinfo {author} {\bibfnamefont {M.-L.}\ \bibnamefont {Bocquet}}, \bibinfo
  {author} {\bibfnamefont {T.}~\bibnamefont {Brugger}}, \bibinfo {author}
  {\bibfnamefont {T.}~\bibnamefont {Greber}}, \bibinfo {author} {\bibfnamefont
  {J.}~\bibnamefont {Wintterlin}}, \ and\ \bibinfo {author} {\bibfnamefont
  {S.}~\bibnamefont {G\"unther}},\ }\href {\doibase
  10.1103/PhysRevLett.104.136102} {\bibfield  {journal} {\bibinfo  {journal}
  {Phys. Rev. Lett.}\ }\textbf {\bibinfo {volume} {104}},\ \bibinfo {pages}
  {136102} (\bibinfo {year} {2010})}\BibitemShut {NoStop}%
\bibitem [{\citenamefont {Hermann}(2012)}]{Hermann2012}%
  \BibitemOpen
  \bibfield  {author} {\bibinfo {author} {\bibfnamefont {K.}~\bibnamefont
  {Hermann}},\ }\href {\doibase 10.1088/0953-8984/24/31/314210} {\bibfield
  {journal} {\bibinfo  {journal} {Journal of Physics: Condensed Matter}\
  }\textbf {\bibinfo {volume} {24}},\ \bibinfo {pages} {314210} (\bibinfo
  {year} {2012})}\BibitemShut {NoStop}%
\bibitem [{\citenamefont {Zeller}\ and\ \citenamefont
  {G\"{u}nther}(2014)}]{Zeller2014}%
  \BibitemOpen
  \bibfield  {author} {\bibinfo {author} {\bibfnamefont {P.}~\bibnamefont
  {Zeller}}\ and\ \bibinfo {author} {\bibfnamefont {S.}~\bibnamefont
  {G\"{u}nther}},\ }\href {http://stacks.iop.org/1367-2630/16/i=8/a=083028}
  {\bibfield  {journal} {\bibinfo  {journal} {New Journal of Physics}\ }\textbf
  {\bibinfo {volume} {16}},\ \bibinfo {pages} {083028} (\bibinfo {year}
  {2014})}\BibitemShut {NoStop}%
\bibitem [{\citenamefont {Moritz}(2015)}]{Moritz2015}%
  \BibitemOpen
  \bibfield  {author} {\bibinfo {author} {\bibfnamefont {W.}~\bibnamefont
  {Moritz}},\ }\href
  {https://www.degruyter.com/view/j/zkri.2015.230.issue-1/zkri-2014-1787/zkri-2014-1787.xml}
  {\bibfield  {journal} {\bibinfo  {journal} {Zeitschrift f\"{u}r
  Kristallographie - Crystalline Materials}\ }\textbf {\bibinfo {volume}
  {230}},\ \bibinfo {pages} {37} (\bibinfo {year} {2015})}\BibitemShut
  {NoStop}%
\bibitem [{\citenamefont {Zeller}\ \emph {et~al.}(2017)\citenamefont {Zeller},
  \citenamefont {Ma},\ and\ \citenamefont {G\"{u}nther}}]{Zeller2017}%
  \BibitemOpen
  \bibfield  {author} {\bibinfo {author} {\bibfnamefont {P.}~\bibnamefont
  {Zeller}}, \bibinfo {author} {\bibfnamefont {X.}~\bibnamefont {Ma}}, \ and\
  \bibinfo {author} {\bibfnamefont {S.}~\bibnamefont {G\"{u}nther}},\ }\href
  {http://stacks.iop.org/1367-2630/19/i=1/a=013015} {\bibfield  {journal}
  {\bibinfo  {journal} {New Journal of Physics}\ }\textbf {\bibinfo {volume}
  {19}},\ \bibinfo {pages} {013015} (\bibinfo {year} {2017})}\BibitemShut
  {NoStop}%
\bibitem [{\citenamefont {Kuznetsov}\ \emph {et~al.}(2018)\citenamefont
  {Kuznetsov}, \citenamefont {Ogorodnikov}, \citenamefont {Usachov},
  \citenamefont {Laubschat}, \citenamefont {Vyalikh}, \citenamefont {Matsui},\
  and\ \citenamefont {Yashina}}]{Kuznetsov2018}%
  \BibitemOpen
  \bibfield  {author} {\bibinfo {author} {\bibfnamefont {M.~V.}\ \bibnamefont
  {Kuznetsov}}, \bibinfo {author} {\bibfnamefont {I.~I.}\ \bibnamefont
  {Ogorodnikov}}, \bibinfo {author} {\bibfnamefont {D.~Y.}\ \bibnamefont
  {Usachov}}, \bibinfo {author} {\bibfnamefont {C.}~\bibnamefont {Laubschat}},
  \bibinfo {author} {\bibfnamefont {D.~V.}\ \bibnamefont {Vyalikh}}, \bibinfo
  {author} {\bibfnamefont {F.}~\bibnamefont {Matsui}}, \ and\ \bibinfo {author}
  {\bibfnamefont {L.~V.}\ \bibnamefont {Yashina}},\ }\href {\doibase
  10.7566/JPSJ.87.061005} {\bibfield  {journal} {\bibinfo  {journal} {Journal
  of the Physical Society of Japan}\ }\textbf {\bibinfo {volume} {87}},\
  \bibinfo {pages} {061005} (\bibinfo {year} {2018})}\BibitemShut {NoStop}%
\bibitem [{\citenamefont {Ster}\ \emph {et~al.}(2019)\citenamefont {Ster},
  \citenamefont {M\"{a}rkl},\ and\ \citenamefont {Brown}}]{Ster2019}%
  \BibitemOpen
  \bibfield  {author} {\bibinfo {author} {\bibfnamefont {M.~L.}\ \bibnamefont
  {Ster}}, \bibinfo {author} {\bibfnamefont {T.}~\bibnamefont {M\"{a}rkl}}, \
  and\ \bibinfo {author} {\bibfnamefont {S.~A.}\ \bibnamefont {Brown}},\ }\href
  {\doibase 10.1088/2053-1583/ab5470} {\bibfield  {journal} {\bibinfo
  {journal} {2D Materials}\ }\textbf {\bibinfo {volume} {7}},\ \bibinfo {pages}
  {011005} (\bibinfo {year} {2019})}\BibitemShut {NoStop}%
\bibitem [{\citenamefont {de~Lima}\ \emph {et~al.}(2020)\citenamefont
  {de~Lima}, \citenamefont {Greber},\ and\ \citenamefont
  {Muntwiler}}]{Lima2020}%
  \BibitemOpen
  \bibfield  {author} {\bibinfo {author} {\bibfnamefont {L.~H.}\ \bibnamefont
  {de~Lima}}, \bibinfo {author} {\bibfnamefont {T.}~\bibnamefont {Greber}}, \
  and\ \bibinfo {author} {\bibfnamefont {M.}~\bibnamefont {Muntwiler}},\ }\href
  {\doibase 10.1088/2053-1583/ab81ae} {\bibfield  {journal} {\bibinfo
  {journal} {2D Materials}\ }\textbf {\bibinfo {volume} {7}},\ \bibinfo {pages}
  {035006} (\bibinfo {year} {2020})}\BibitemShut {NoStop}%
\bibitem [{\citenamefont {Muntwiler}\ \emph {et~al.}(2001)\citenamefont
  {Muntwiler}, \citenamefont {Auwärter}, \citenamefont {Baumberger},
  \citenamefont {Hoesch}, \citenamefont {Greber},\ and\ \citenamefont
  {Osterwalder}}]{Muntwiler2001}%
  \BibitemOpen
  \bibfield  {author} {\bibinfo {author} {\bibfnamefont {M.}~\bibnamefont
  {Muntwiler}}, \bibinfo {author} {\bibfnamefont {W.}~\bibnamefont
  {Auwärter}}, \bibinfo {author} {\bibfnamefont {F.}~\bibnamefont
  {Baumberger}}, \bibinfo {author} {\bibfnamefont {M.}~\bibnamefont {Hoesch}},
  \bibinfo {author} {\bibfnamefont {T.}~\bibnamefont {Greber}}, \ and\ \bibinfo
  {author} {\bibfnamefont {J.}~\bibnamefont {Osterwalder}},\ }\href {\doibase
  https://doi.org/10.1016/S0039-6028(00)00928-6} {\bibfield  {journal}
  {\bibinfo  {journal} {Surface Science}\ }\textbf {\bibinfo {volume} {472}},\
  \bibinfo {pages} {125 } (\bibinfo {year} {2001})}\BibitemShut {NoStop}%
\bibitem [{\citenamefont {Rehr}\ and\ \citenamefont {Albers}(1990)}]{Rehr1990}%
  \BibitemOpen
  \bibfield  {author} {\bibinfo {author} {\bibfnamefont {J.~J.}\ \bibnamefont
  {Rehr}}\ and\ \bibinfo {author} {\bibfnamefont {R.~C.}\ \bibnamefont
  {Albers}},\ }\href {\doibase 10.1103/PhysRevB.41.8139} {\bibfield  {journal}
  {\bibinfo  {journal} {Phys. Rev. B}\ }\textbf {\bibinfo {volume} {41}},\
  \bibinfo {pages} {8139} (\bibinfo {year} {1990})}\BibitemShut {NoStop}%
\bibitem [{\citenamefont {Chen}\ \emph {et~al.}(1998)\citenamefont {Chen},
  \citenamefont {Garc\'{\i}a~de Abajo}, \citenamefont {Chass\'e}, \citenamefont
  {Ynzunza}, \citenamefont {Kaduwela}, \citenamefont {Van~Hove},\ and\
  \citenamefont {Fadley}}]{Chen1998}%
  \BibitemOpen
  \bibfield  {author} {\bibinfo {author} {\bibfnamefont {Y.}~\bibnamefont
  {Chen}}, \bibinfo {author} {\bibfnamefont {F.~J.}\ \bibnamefont
  {Garc\'{\i}a~de Abajo}}, \bibinfo {author} {\bibfnamefont {A.}~\bibnamefont
  {Chass\'e}}, \bibinfo {author} {\bibfnamefont {R.~X.}\ \bibnamefont
  {Ynzunza}}, \bibinfo {author} {\bibfnamefont {A.~P.}\ \bibnamefont
  {Kaduwela}}, \bibinfo {author} {\bibfnamefont {M.~A.}\ \bibnamefont
  {Van~Hove}}, \ and\ \bibinfo {author} {\bibfnamefont {C.~S.}\ \bibnamefont
  {Fadley}},\ }\href {\doibase 10.1103/PhysRevB.58.13121} {\bibfield  {journal}
  {\bibinfo  {journal} {Phys. Rev. B}\ }\textbf {\bibinfo {volume} {58}},\
  \bibinfo {pages} {13121} (\bibinfo {year} {1998})}\BibitemShut {NoStop}%
\bibitem [{\citenamefont {de~Siervo}\ \emph {et~al.}(2002)\citenamefont
  {de~Siervo}, \citenamefont {Soares}, \citenamefont {Landers}, \citenamefont
  {Fazan}, \citenamefont {Morais},\ and\ \citenamefont {Kleiman}}]{Siervo2002}%
  \BibitemOpen
  \bibfield  {author} {\bibinfo {author} {\bibfnamefont {A.}~\bibnamefont
  {de~Siervo}}, \bibinfo {author} {\bibfnamefont {E.}~\bibnamefont {Soares}},
  \bibinfo {author} {\bibfnamefont {R.}~\bibnamefont {Landers}}, \bibinfo
  {author} {\bibfnamefont {T.~A.}\ \bibnamefont {Fazan}}, \bibinfo {author}
  {\bibfnamefont {J.}~\bibnamefont {Morais}}, \ and\ \bibinfo {author}
  {\bibfnamefont {G.}~\bibnamefont {Kleiman}},\ }\href {\doibase
  http://dx.doi.org/10.1016/S0039-6028(02)01097-X} {\bibfield  {journal}
  {\bibinfo  {journal} {Surface Science}\ }\textbf {\bibinfo {volume} {504}},\
  \bibinfo {pages} {215 } (\bibinfo {year} {2002})}\BibitemShut {NoStop}%
\bibitem [{\citenamefont {Soares}\ \emph {et~al.}(2002)\citenamefont {Soares},
  \citenamefont {de~Siervo}, \citenamefont {Landers},\ and\ \citenamefont
  {Kleiman}}]{Soares2002}%
  \BibitemOpen
  \bibfield  {author} {\bibinfo {author} {\bibfnamefont {E.}~\bibnamefont
  {Soares}}, \bibinfo {author} {\bibfnamefont {A.}~\bibnamefont {de~Siervo}},
  \bibinfo {author} {\bibfnamefont {R.}~\bibnamefont {Landers}}, \ and\
  \bibinfo {author} {\bibfnamefont {G.}~\bibnamefont {Kleiman}},\ }\href
  {\doibase https://doi.org/10.1016/S0039-6028(01)01561-8} {\bibfield
  {journal} {\bibinfo  {journal} {Surface Science}\ }\textbf {\bibinfo {volume}
  {497}},\ \bibinfo {pages} {205 } (\bibinfo {year} {2002})}\BibitemShut
  {NoStop}%
\bibitem [{\citenamefont {Cezar}\ \emph {et~al.}(2013)\citenamefont {Cezar},
  \citenamefont {Fonseca}, \citenamefont {Rodrigues}, \citenamefont
  {de~Castro}, \citenamefont {Neuenschwander}, \citenamefont {Rodrigues},
  \citenamefont {Meyer}, \citenamefont {Ribeiro}, \citenamefont {Moreira},
  \citenamefont {Piton}, \citenamefont {Raulik}, \citenamefont {Donadio},
  \citenamefont {Seraphim}, \citenamefont {Barbosa}, \citenamefont {de~Siervo},
  \citenamefont {Landers},\ and\ \citenamefont {de~Brito}}]{Cezar2013}%
  \BibitemOpen
  \bibfield  {author} {\bibinfo {author} {\bibfnamefont {J.~C.}\ \bibnamefont
  {Cezar}}, \bibinfo {author} {\bibfnamefont {P.~T.}\ \bibnamefont {Fonseca}},
  \bibinfo {author} {\bibfnamefont {G.~L. M.~P.}\ \bibnamefont {Rodrigues}},
  \bibinfo {author} {\bibfnamefont {A.~R.~B.}\ \bibnamefont {de~Castro}},
  \bibinfo {author} {\bibfnamefont {R.~T.}\ \bibnamefont {Neuenschwander}},
  \bibinfo {author} {\bibfnamefont {F.}~\bibnamefont {Rodrigues}}, \bibinfo
  {author} {\bibfnamefont {B.~C.}\ \bibnamefont {Meyer}}, \bibinfo {author}
  {\bibfnamefont {L.~F.~S.}\ \bibnamefont {Ribeiro}}, \bibinfo {author}
  {\bibfnamefont {A.~F. A.~G.}\ \bibnamefont {Moreira}}, \bibinfo {author}
  {\bibfnamefont {J.~R.}\ \bibnamefont {Piton}}, \bibinfo {author}
  {\bibfnamefont {M.~A.}\ \bibnamefont {Raulik}}, \bibinfo {author}
  {\bibfnamefont {M.~P.}\ \bibnamefont {Donadio}}, \bibinfo {author}
  {\bibfnamefont {R.~M.}\ \bibnamefont {Seraphim}}, \bibinfo {author}
  {\bibfnamefont {M.~A.}\ \bibnamefont {Barbosa}}, \bibinfo {author}
  {\bibfnamefont {A.}~\bibnamefont {de~Siervo}}, \bibinfo {author}
  {\bibfnamefont {R.}~\bibnamefont {Landers}}, \ and\ \bibinfo {author}
  {\bibfnamefont {A.~N.}\ \bibnamefont {de~Brito}},\ }\href
  {http://stacks.iop.org/1742-6596/425/i=7/a=072015} {\bibfield  {journal}
  {\bibinfo  {journal} {Journal of Physics: Conference Series}\ }\textbf
  {\bibinfo {volume} {425}},\ \bibinfo {pages} {072015} (\bibinfo {year}
  {2013})}\BibitemShut {NoStop}%
\bibitem [{\citenamefont {Booth}\ \emph {et~al.}(1997)\citenamefont {Booth},
  \citenamefont {Davis}, \citenamefont {Toomes}, \citenamefont {Woodruff},
  \citenamefont {Hirschmugl}, \citenamefont {Schindler}, \citenamefont
  {Schaff}, \citenamefont {Fernandez}, \citenamefont {Theobald}, \citenamefont
  {Hofmann}, \citenamefont {Lindsay}, \citenamefont {Gießel}, \citenamefont
  {Baumgärtel},\ and\ \citenamefont {Bradshaw}}]{Booth1997}%
  \BibitemOpen
  \bibfield  {author} {\bibinfo {author} {\bibfnamefont {N.}~\bibnamefont
  {Booth}}, \bibinfo {author} {\bibfnamefont {R.}~\bibnamefont {Davis}},
  \bibinfo {author} {\bibfnamefont {R.}~\bibnamefont {Toomes}}, \bibinfo
  {author} {\bibfnamefont {D.}~\bibnamefont {Woodruff}}, \bibinfo {author}
  {\bibfnamefont {C.}~\bibnamefont {Hirschmugl}}, \bibinfo {author}
  {\bibfnamefont {K.}~\bibnamefont {Schindler}}, \bibinfo {author}
  {\bibfnamefont {O.}~\bibnamefont {Schaff}}, \bibinfo {author} {\bibfnamefont
  {V.}~\bibnamefont {Fernandez}}, \bibinfo {author} {\bibfnamefont
  {A.}~\bibnamefont {Theobald}}, \bibinfo {author} {\bibfnamefont
  {P.}~\bibnamefont {Hofmann}}, \bibinfo {author} {\bibfnamefont
  {R.}~\bibnamefont {Lindsay}}, \bibinfo {author} {\bibfnamefont
  {T.}~\bibnamefont {Gießel}}, \bibinfo {author} {\bibfnamefont
  {P.}~\bibnamefont {Baumgärtel}}, \ and\ \bibinfo {author} {\bibfnamefont
  {A.}~\bibnamefont {Bradshaw}},\ }\href {\doibase
  https://doi.org/10.1016/S0039-6028(97)00280-X} {\bibfield  {journal}
  {\bibinfo  {journal} {Surface Science}\ }\textbf {\bibinfo {volume} {387}},\
  \bibinfo {pages} {152 } (\bibinfo {year} {1997})}\BibitemShut {NoStop}%
\bibitem [{\citenamefont {Pendry}(1980)}]{Pendry1980}%
  \BibitemOpen
  \bibfield  {author} {\bibinfo {author} {\bibfnamefont {J.~B.}\ \bibnamefont
  {Pendry}},\ }\href {\doibase 10.1088/0022-3719/13/5/024} {\bibfield
  {journal} {\bibinfo  {journal} {Journal of Physics C: Solid State Physics}\
  }\textbf {\bibinfo {volume} {13}},\ \bibinfo {pages} {937} (\bibinfo {year}
  {1980})}\BibitemShut {NoStop}%
\bibitem [{\citenamefont {Bondino}\ \emph {et~al.}(2002)\citenamefont
  {Bondino}, \citenamefont {Comelli}, \citenamefont {Baraldi}, \citenamefont
  {Rosei}, \citenamefont {Lizzit}, \citenamefont {Goldoni}, \citenamefont
  {Larciprete},\ and\ \citenamefont {Paolucci}}]{Bondino2002}%
  \BibitemOpen
  \bibfield  {author} {\bibinfo {author} {\bibfnamefont {F.}~\bibnamefont
  {Bondino}}, \bibinfo {author} {\bibfnamefont {G.}~\bibnamefont {Comelli}},
  \bibinfo {author} {\bibfnamefont {A.}~\bibnamefont {Baraldi}}, \bibinfo
  {author} {\bibfnamefont {R.}~\bibnamefont {Rosei}}, \bibinfo {author}
  {\bibfnamefont {S.}~\bibnamefont {Lizzit}}, \bibinfo {author} {\bibfnamefont
  {A.}~\bibnamefont {Goldoni}}, \bibinfo {author} {\bibfnamefont
  {R.}~\bibnamefont {Larciprete}}, \ and\ \bibinfo {author} {\bibfnamefont
  {G.}~\bibnamefont {Paolucci}},\ }\href {\doibase 10.1103/PhysRevB.66.075402}
  {\bibfield  {journal} {\bibinfo  {journal} {Phys. Rev. B}\ }\textbf {\bibinfo
  {volume} {66}},\ \bibinfo {pages} {075402} (\bibinfo {year}
  {2002})}\BibitemShut {NoStop}%
\bibitem [{\citenamefont {van~der Veen}\ \emph {et~al.}(1980)\citenamefont
  {van~der Veen}, \citenamefont {Himpsel},\ and\ \citenamefont
  {Eastman}}]{Veen1980}%
  \BibitemOpen
  \bibfield  {author} {\bibinfo {author} {\bibfnamefont {J.~F.}\ \bibnamefont
  {van~der Veen}}, \bibinfo {author} {\bibfnamefont {F.~J.}\ \bibnamefont
  {Himpsel}}, \ and\ \bibinfo {author} {\bibfnamefont {D.~E.}\ \bibnamefont
  {Eastman}},\ }\href {\doibase 10.1103/PhysRevLett.44.189} {\bibfield
  {journal} {\bibinfo  {journal} {Phys. Rev. Lett.}\ }\textbf {\bibinfo
  {volume} {44}},\ \bibinfo {pages} {189} (\bibinfo {year} {1980})}\BibitemShut
  {NoStop}%
\bibitem [{\citenamefont {Bianchi}\ \emph {et~al.}(2009)\citenamefont
  {Bianchi}, \citenamefont {Cassese}, \citenamefont {Cavallin}, \citenamefont
  {Comin}, \citenamefont {Orlando}, \citenamefont {Postregna}, \citenamefont
  {Golfetto}, \citenamefont {Lizzit},\ and\ \citenamefont
  {Baraldi}}]{Bianchi2009}%
  \BibitemOpen
  \bibfield  {author} {\bibinfo {author} {\bibfnamefont {M.}~\bibnamefont
  {Bianchi}}, \bibinfo {author} {\bibfnamefont {D.}~\bibnamefont {Cassese}},
  \bibinfo {author} {\bibfnamefont {A.}~\bibnamefont {Cavallin}}, \bibinfo
  {author} {\bibfnamefont {R.}~\bibnamefont {Comin}}, \bibinfo {author}
  {\bibfnamefont {F.}~\bibnamefont {Orlando}}, \bibinfo {author} {\bibfnamefont
  {L.}~\bibnamefont {Postregna}}, \bibinfo {author} {\bibfnamefont
  {E.}~\bibnamefont {Golfetto}}, \bibinfo {author} {\bibfnamefont
  {S.}~\bibnamefont {Lizzit}}, \ and\ \bibinfo {author} {\bibfnamefont
  {A.}~\bibnamefont {Baraldi}},\ }\href
  {http://stacks.iop.org/1367-2630/11/i=6/a=063002} {\bibfield  {journal}
  {\bibinfo  {journal} {New Journal of Physics}\ }\textbf {\bibinfo {volume}
  {11}},\ \bibinfo {pages} {063002} (\bibinfo {year} {2009})}\BibitemShut
  {NoStop}%
\bibitem [{\citenamefont {Lacovig}\ \emph {et~al.}(2009)\citenamefont
  {Lacovig}, \citenamefont {Pozzo}, \citenamefont {Alf\`e}, \citenamefont
  {Vilmercati}, \citenamefont {Baraldi},\ and\ \citenamefont
  {Lizzit}}]{Lacovig2009}%
  \BibitemOpen
  \bibfield  {author} {\bibinfo {author} {\bibfnamefont {P.}~\bibnamefont
  {Lacovig}}, \bibinfo {author} {\bibfnamefont {M.}~\bibnamefont {Pozzo}},
  \bibinfo {author} {\bibfnamefont {D.}~\bibnamefont {Alf\`e}}, \bibinfo
  {author} {\bibfnamefont {P.}~\bibnamefont {Vilmercati}}, \bibinfo {author}
  {\bibfnamefont {A.}~\bibnamefont {Baraldi}}, \ and\ \bibinfo {author}
  {\bibfnamefont {S.}~\bibnamefont {Lizzit}},\ }\href {\doibase
  10.1103/PhysRevLett.103.166101} {\bibfield  {journal} {\bibinfo  {journal}
  {Phys. Rev. Lett.}\ }\textbf {\bibinfo {volume} {103}},\ \bibinfo {pages}
  {166101} (\bibinfo {year} {2009})}\BibitemShut {NoStop}%
\bibitem [{\citenamefont {Tanuma}\ \emph {et~al.}(1991)\citenamefont {Tanuma},
  \citenamefont {Powell},\ and\ \citenamefont {Penn}}]{Tanuma1991}%
  \BibitemOpen
  \bibfield  {author} {\bibinfo {author} {\bibfnamefont {S.}~\bibnamefont
  {Tanuma}}, \bibinfo {author} {\bibfnamefont {C.~J.}\ \bibnamefont {Powell}},
  \ and\ \bibinfo {author} {\bibfnamefont {D.~R.}\ \bibnamefont {Penn}},\
  }\href {\doibase 10.1002/sia.740171304} {\bibfield  {journal} {\bibinfo
  {journal} {Surface and Interface Analysis}\ }\textbf {\bibinfo {volume}
  {17}},\ \bibinfo {pages} {911} (\bibinfo {year} {1991})}\BibitemShut
  {NoStop}%
\bibitem [{\citenamefont {Powell}\ and\ \citenamefont
  {Jablonski}(2010)}]{Powell2010}%
  \BibitemOpen
  \bibfield  {author} {\bibinfo {author} {\bibfnamefont {C.~J.}\ \bibnamefont
  {Powell}}\ and\ \bibinfo {author} {\bibfnamefont {A.}~\bibnamefont
  {Jablonski}},\ }\href@noop {} {\emph {\bibinfo {title} {NIST Electron
  Inelastic-Mean-Free-Path Database - Version 1.2}}}\ (\bibinfo  {publisher}
  {National Institute of Standards and Technology, Gaithersburg, MD},\ \bibinfo
  {year} {2010})\BibitemShut {NoStop}%
\bibitem [{\citenamefont {de~Lima}\ \emph {et~al.}(2013)\citenamefont
  {de~Lima}, \citenamefont {de~Siervo}, \citenamefont {Landers}, \citenamefont
  {Viana}, \citenamefont {Goncalves}, \citenamefont {Lacerda},\ and\
  \citenamefont {H\"aberle}}]{Lima2013}%
  \BibitemOpen
  \bibfield  {author} {\bibinfo {author} {\bibfnamefont {L.~H.}\ \bibnamefont
  {de~Lima}}, \bibinfo {author} {\bibfnamefont {A.}~\bibnamefont {de~Siervo}},
  \bibinfo {author} {\bibfnamefont {R.}~\bibnamefont {Landers}}, \bibinfo
  {author} {\bibfnamefont {G.~A.}\ \bibnamefont {Viana}}, \bibinfo {author}
  {\bibfnamefont {A.~M.~B.}\ \bibnamefont {Goncalves}}, \bibinfo {author}
  {\bibfnamefont {R.~G.}\ \bibnamefont {Lacerda}}, \ and\ \bibinfo {author}
  {\bibfnamefont {P.}~\bibnamefont {H\"aberle}},\ }\href {\doibase
  10.1103/PhysRevB.87.081403} {\bibfield  {journal} {\bibinfo  {journal} {Phys.
  Rev. B}\ }\textbf {\bibinfo {volume} {87}},\ \bibinfo {pages} {081403}
  (\bibinfo {year} {2013})}\BibitemShut {NoStop}%
\bibitem [{\citenamefont {Ulstrup}\ \emph {et~al.}(2018)\citenamefont
  {Ulstrup}, \citenamefont {Lacovig}, \citenamefont {Orlando}, \citenamefont
  {Lizzit}, \citenamefont {Bignardi}, \citenamefont {Dalmiglio}, \citenamefont
  {Bianchi}, \citenamefont {Mazzola}, \citenamefont {Baraldi}, \citenamefont
  {Larciprete}, \citenamefont {Hofmann},\ and\ \citenamefont
  {Lizzit}}]{Ulstrup2018}%
  \BibitemOpen
  \bibfield  {author} {\bibinfo {author} {\bibfnamefont {S.}~\bibnamefont
  {Ulstrup}}, \bibinfo {author} {\bibfnamefont {P.}~\bibnamefont {Lacovig}},
  \bibinfo {author} {\bibfnamefont {F.}~\bibnamefont {Orlando}}, \bibinfo
  {author} {\bibfnamefont {D.}~\bibnamefont {Lizzit}}, \bibinfo {author}
  {\bibfnamefont {L.}~\bibnamefont {Bignardi}}, \bibinfo {author}
  {\bibfnamefont {M.}~\bibnamefont {Dalmiglio}}, \bibinfo {author}
  {\bibfnamefont {M.}~\bibnamefont {Bianchi}}, \bibinfo {author} {\bibfnamefont
  {F.}~\bibnamefont {Mazzola}}, \bibinfo {author} {\bibfnamefont
  {A.}~\bibnamefont {Baraldi}}, \bibinfo {author} {\bibfnamefont
  {R.}~\bibnamefont {Larciprete}}, \bibinfo {author} {\bibfnamefont
  {P.}~\bibnamefont {Hofmann}}, \ and\ \bibinfo {author} {\bibfnamefont
  {S.}~\bibnamefont {Lizzit}},\ }\href {\doibase
  https://doi.org/10.1016/j.susc.2018.03.017} {\bibfield  {journal} {\bibinfo
  {journal} {Surface Science}\ }\textbf {\bibinfo {volume} {678}},\ \bibinfo
  {pages} {57 } (\bibinfo {year} {2018})}\BibitemShut {NoStop}%
\bibitem [{\citenamefont {Westphal}(2003)}]{Westphal2003}%
  \BibitemOpen
  \bibfield  {author} {\bibinfo {author} {\bibfnamefont {C.}~\bibnamefont
  {Westphal}},\ }\href {\doibase
  http://dx.doi.org/10.1016/S0167-5729(03)00022-0} {\bibfield  {journal}
  {\bibinfo  {journal} {Surface Science Reports}\ }\textbf {\bibinfo {volume}
  {50}},\ \bibinfo {pages} {1 } (\bibinfo {year} {2003})}\BibitemShut {NoStop}%
\bibitem [{\citenamefont {de~Lima}\ \emph
  {et~al.}(2014{\natexlab{a}})\citenamefont {de~Lima}, \citenamefont
  {Landers},\ and\ \citenamefont {de~Siervo}}]{Lima2014}%
  \BibitemOpen
  \bibfield  {author} {\bibinfo {author} {\bibfnamefont {L.~H.}\ \bibnamefont
  {de~Lima}}, \bibinfo {author} {\bibfnamefont {R.}~\bibnamefont {Landers}}, \
  and\ \bibinfo {author} {\bibfnamefont {A.}~\bibnamefont {de~Siervo}},\ }\href
  {\doibase 10.1021/cm501976b} {\bibfield  {journal} {\bibinfo  {journal}
  {Chemistry of Materials}\ }\textbf {\bibinfo {volume} {26}},\ \bibinfo
  {pages} {4172} (\bibinfo {year} {2014}{\natexlab{a}})}\BibitemShut {NoStop}%
\bibitem [{\citenamefont {de~Lima}\ \emph
  {et~al.}(2014{\natexlab{b}})\citenamefont {de~Lima}, \citenamefont
  {Handschak}, \citenamefont {Schönbohm}, \citenamefont {Landers},
  \citenamefont {Westphal},\ and\ \citenamefont {de~Siervo}}]{Lima2014a}%
  \BibitemOpen
  \bibfield  {author} {\bibinfo {author} {\bibfnamefont {L.~H.}\ \bibnamefont
  {de~Lima}}, \bibinfo {author} {\bibfnamefont {D.}~\bibnamefont {Handschak}},
  \bibinfo {author} {\bibfnamefont {F.}~\bibnamefont {Schönbohm}}, \bibinfo
  {author} {\bibfnamefont {R.}~\bibnamefont {Landers}}, \bibinfo {author}
  {\bibfnamefont {C.}~\bibnamefont {Westphal}}, \ and\ \bibinfo {author}
  {\bibfnamefont {A.}~\bibnamefont {de~Siervo}},\ }\href {\doibase
  10.1039/C4CC05005C} {\bibfield  {journal} {\bibinfo  {journal} {Chem.
  Commun.}\ }\textbf {\bibinfo {volume} {50}},\ \bibinfo {pages} {13571}
  (\bibinfo {year} {2014}{\natexlab{b}})}\BibitemShut {NoStop}%
\bibitem [{\citenamefont {de~Lima}\ \emph {et~al.}(2016)\citenamefont
  {de~Lima}, \citenamefont {Barreto}, \citenamefont {Landers},\ and\
  \citenamefont {de~Siervo}}]{Lima2016}%
  \BibitemOpen
  \bibfield  {author} {\bibinfo {author} {\bibfnamefont {L.~H.}\ \bibnamefont
  {de~Lima}}, \bibinfo {author} {\bibfnamefont {L.}~\bibnamefont {Barreto}},
  \bibinfo {author} {\bibfnamefont {R.}~\bibnamefont {Landers}}, \ and\
  \bibinfo {author} {\bibfnamefont {A.}~\bibnamefont {de~Siervo}},\ }\href
  {\doibase 10.1103/PhysRevB.93.035448} {\bibfield  {journal} {\bibinfo
  {journal} {Phys. Rev. B}\ }\textbf {\bibinfo {volume} {93}},\ \bibinfo
  {pages} {035448} (\bibinfo {year} {2016})}\BibitemShut {NoStop}%
\end{thebibliography}%


\begin{thebibliography}{9}%
\makeatletter
\providecommand \@ifxundefined [1]{%
 \@ifx{#1\undefined}
}%
\providecommand \@ifnum [1]{%
 \ifnum #1\expandafter \@firstoftwo
 \else \expandafter \@secondoftwo
 \fi
}%
\providecommand \@ifx [1]{%
 \ifx #1\expandafter \@firstoftwo
 \else \expandafter \@secondoftwo
 \fi
}%
\providecommand \natexlab [1]{#1}%
\providecommand \enquote  [1]{``#1''}%
\providecommand \bibnamefont  [1]{#1}%
\providecommand \bibfnamefont [1]{#1}%
\providecommand \citenamefont [1]{#1}%
\providecommand \href@noop [0]{\@secondoftwo}%
\providecommand \href [0]{\begingroup \@sanitize@url \@href}%
\providecommand \@href[1]{\@@startlink{#1}\@@href}%
\providecommand \@@href[1]{\endgroup#1\@@endlink}%
\providecommand \@sanitize@url [0]{\catcode `\\12\catcode `\$12\catcode
  `\&12\catcode `\#12\catcode `\^12\catcode `\_12\catcode `\%12\relax}%
\providecommand \@@startlink[1]{}%
\providecommand \@@endlink[0]{}%
\providecommand \url  [0]{\begingroup\@sanitize@url \@url }%
\providecommand \@url [1]{\endgroup\@href {#1}{\urlprefix }}%
\providecommand \urlprefix  [0]{URL }%
\providecommand \Eprint [0]{\href }%
\providecommand \doibase [0]{http://dx.doi.org/}%
\providecommand \selectlanguage [0]{\@gobble}%
\providecommand \bibinfo  [0]{\@secondoftwo}%
\providecommand \bibfield  [0]{\@secondoftwo}%
\providecommand \translation [1]{[#1]}%
\providecommand \BibitemOpen [0]{}%
\providecommand \bibitemStop [0]{}%
\providecommand \bibitemNoStop [0]{.\EOS\space}%
\providecommand \EOS [0]{\spacefactor3000\relax}%
\providecommand \BibitemShut  [1]{\csname bibitem#1\endcsname}%
\let\auto@bib@innerbib\@empty
\bibitem [{\citenamefont {Cezar}\ \emph {et~al.}(2013)\citenamefont {Cezar},
  \citenamefont {Fonseca}, \citenamefont {Rodrigues}, \citenamefont
  {de~Castro}, \citenamefont {Neuenschwander}, \citenamefont {Rodrigues},
  \citenamefont {Meyer}, \citenamefont {Ribeiro}, \citenamefont {Moreira},
  \citenamefont {Piton}, \citenamefont {Raulik}, \citenamefont {Donadio},
  \citenamefont {Seraphim}, \citenamefont {Barbosa}, \citenamefont {de~Siervo},
  \citenamefont {Landers},\ and\ \citenamefont {de~Brito}}]{Cezar2013}%
  \BibitemOpen
  \bibfield  {author} {\bibinfo {author} {\bibfnamefont {J.~C.}\ \bibnamefont
  {Cezar}}, \bibinfo {author} {\bibfnamefont {P.~T.}\ \bibnamefont {Fonseca}},
  \bibinfo {author} {\bibfnamefont {G.~L. M.~P.}\ \bibnamefont {Rodrigues}},
  \bibinfo {author} {\bibfnamefont {A.~R.~B.}\ \bibnamefont {de~Castro}},
  \bibinfo {author} {\bibfnamefont {R.~T.}\ \bibnamefont {Neuenschwander}},
  \bibinfo {author} {\bibfnamefont {F.}~\bibnamefont {Rodrigues}}, \bibinfo
  {author} {\bibfnamefont {B.~C.}\ \bibnamefont {Meyer}}, \bibinfo {author}
  {\bibfnamefont {L.~F.~S.}\ \bibnamefont {Ribeiro}}, \bibinfo {author}
  {\bibfnamefont {A.~F. A.~G.}\ \bibnamefont {Moreira}}, \bibinfo {author}
  {\bibfnamefont {J.~R.}\ \bibnamefont {Piton}}, \bibinfo {author}
  {\bibfnamefont {M.~A.}\ \bibnamefont {Raulik}}, \bibinfo {author}
  {\bibfnamefont {M.~P.}\ \bibnamefont {Donadio}}, \bibinfo {author}
  {\bibfnamefont {R.~M.}\ \bibnamefont {Seraphim}}, \bibinfo {author}
  {\bibfnamefont {M.~A.}\ \bibnamefont {Barbosa}}, \bibinfo {author}
  {\bibfnamefont {A.}~\bibnamefont {de~Siervo}}, \bibinfo {author}
  {\bibfnamefont {R.}~\bibnamefont {Landers}}, \ and\ \bibinfo {author}
  {\bibfnamefont {A.~N.}\ \bibnamefont {de~Brito}},\ }\href
  {http://stacks.iop.org/1742-6596/425/i=7/a=072015} {\bibfield  {journal}
  {\bibinfo  {journal} {Journal of Physics: Conference Series}\ }\textbf
  {\bibinfo {volume} {425}},\ \bibinfo {pages} {072015} (\bibinfo {year}
  {2013})}\BibitemShut {NoStop}%
\bibitem [{\citenamefont {Rehr}\ and\ \citenamefont {Albers}(1990)}]{Rehr1990}%
  \BibitemOpen
  \bibfield  {author} {\bibinfo {author} {\bibfnamefont {J.~J.}\ \bibnamefont
  {Rehr}}\ and\ \bibinfo {author} {\bibfnamefont {R.~C.}\ \bibnamefont
  {Albers}},\ }\href {\doibase 10.1103/PhysRevB.41.8139} {\bibfield  {journal}
  {\bibinfo  {journal} {Phys. Rev. B}\ }\textbf {\bibinfo {volume} {41}},\
  \bibinfo {pages} {8139} (\bibinfo {year} {1990})}\BibitemShut {NoStop}%
\bibitem [{\citenamefont {Chen}\ \emph {et~al.}(1998)\citenamefont {Chen},
  \citenamefont {Garc\'{\i}a~de Abajo}, \citenamefont {Chass\'e}, \citenamefont
  {Ynzunza}, \citenamefont {Kaduwela}, \citenamefont {Van~Hove},\ and\
  \citenamefont {Fadley}}]{Chen1998}%
  \BibitemOpen
  \bibfield  {author} {\bibinfo {author} {\bibfnamefont {Y.}~\bibnamefont
  {Chen}}, \bibinfo {author} {\bibfnamefont {F.~J.}\ \bibnamefont
  {Garc\'{\i}a~de Abajo}}, \bibinfo {author} {\bibfnamefont {A.}~\bibnamefont
  {Chass\'e}}, \bibinfo {author} {\bibfnamefont {R.~X.}\ \bibnamefont
  {Ynzunza}}, \bibinfo {author} {\bibfnamefont {A.~P.}\ \bibnamefont
  {Kaduwela}}, \bibinfo {author} {\bibfnamefont {M.~A.}\ \bibnamefont
  {Van~Hove}}, \ and\ \bibinfo {author} {\bibfnamefont {C.~S.}\ \bibnamefont
  {Fadley}},\ }\href {\doibase 10.1103/PhysRevB.58.13121} {\bibfield  {journal}
  {\bibinfo  {journal} {Phys. Rev. B}\ }\textbf {\bibinfo {volume} {58}},\
  \bibinfo {pages} {13121} (\bibinfo {year} {1998})}\BibitemShut {NoStop}%
\bibitem [{\citenamefont {de~Siervo}\ \emph {et~al.}(2002)\citenamefont
  {de~Siervo}, \citenamefont {Soares}, \citenamefont {Landers}, \citenamefont
  {Fazan}, \citenamefont {Morais},\ and\ \citenamefont {Kleiman}}]{Siervo2002}%
  \BibitemOpen
  \bibfield  {author} {\bibinfo {author} {\bibfnamefont {A.}~\bibnamefont
  {de~Siervo}}, \bibinfo {author} {\bibfnamefont {E.}~\bibnamefont {Soares}},
  \bibinfo {author} {\bibfnamefont {R.}~\bibnamefont {Landers}}, \bibinfo
  {author} {\bibfnamefont {T.~A.}\ \bibnamefont {Fazan}}, \bibinfo {author}
  {\bibfnamefont {J.}~\bibnamefont {Morais}}, \ and\ \bibinfo {author}
  {\bibfnamefont {G.}~\bibnamefont {Kleiman}},\ }\href {\doibase
  http://dx.doi.org/10.1016/S0039-6028(02)01097-X} {\bibfield  {journal}
  {\bibinfo  {journal} {Surface Science}\ }\textbf {\bibinfo {volume} {504}},\
  \bibinfo {pages} {215 } (\bibinfo {year} {2002})}\BibitemShut {NoStop}%
\bibitem [{\citenamefont {Soares}\ \emph {et~al.}(2002)\citenamefont {Soares},
  \citenamefont {de~Siervo}, \citenamefont {Landers},\ and\ \citenamefont
  {Kleiman}}]{Soares2002}%
  \BibitemOpen
  \bibfield  {author} {\bibinfo {author} {\bibfnamefont {E.}~\bibnamefont
  {Soares}}, \bibinfo {author} {\bibfnamefont {A.}~\bibnamefont {de~Siervo}},
  \bibinfo {author} {\bibfnamefont {R.}~\bibnamefont {Landers}}, \ and\
  \bibinfo {author} {\bibfnamefont {G.}~\bibnamefont {Kleiman}},\ }\href
  {\doibase https://doi.org/10.1016/S0039-6028(01)01561-8} {\bibfield
  {journal} {\bibinfo  {journal} {Surface Science}\ }\textbf {\bibinfo {volume}
  {497}},\ \bibinfo {pages} {205 } (\bibinfo {year} {2002})}\BibitemShut
  {NoStop}%
\bibitem [{\citenamefont {Booth}\ \emph {et~al.}(1997)\citenamefont {Booth},
  \citenamefont {Davis}, \citenamefont {Toomes}, \citenamefont {Woodruff},
  \citenamefont {Hirschmugl}, \citenamefont {Schindler}, \citenamefont
  {Schaff}, \citenamefont {Fernandez}, \citenamefont {Theobald}, \citenamefont
  {Hofmann}, \citenamefont {Lindsay}, \citenamefont {Gießel}, \citenamefont
  {Baumgärtel},\ and\ \citenamefont {Bradshaw}}]{Booth1997}%
  \BibitemOpen
  \bibfield  {author} {\bibinfo {author} {\bibfnamefont {N.}~\bibnamefont
  {Booth}}, \bibinfo {author} {\bibfnamefont {R.}~\bibnamefont {Davis}},
  \bibinfo {author} {\bibfnamefont {R.}~\bibnamefont {Toomes}}, \bibinfo
  {author} {\bibfnamefont {D.}~\bibnamefont {Woodruff}}, \bibinfo {author}
  {\bibfnamefont {C.}~\bibnamefont {Hirschmugl}}, \bibinfo {author}
  {\bibfnamefont {K.}~\bibnamefont {Schindler}}, \bibinfo {author}
  {\bibfnamefont {O.}~\bibnamefont {Schaff}}, \bibinfo {author} {\bibfnamefont
  {V.}~\bibnamefont {Fernandez}}, \bibinfo {author} {\bibfnamefont
  {A.}~\bibnamefont {Theobald}}, \bibinfo {author} {\bibfnamefont
  {P.}~\bibnamefont {Hofmann}}, \bibinfo {author} {\bibfnamefont
  {R.}~\bibnamefont {Lindsay}}, \bibinfo {author} {\bibfnamefont
  {T.}~\bibnamefont {Gießel}}, \bibinfo {author} {\bibfnamefont
  {P.}~\bibnamefont {Baumgärtel}}, \ and\ \bibinfo {author} {\bibfnamefont
  {A.}~\bibnamefont {Bradshaw}},\ }\href {\doibase
  https://doi.org/10.1016/S0039-6028(97)00280-X} {\bibfield  {journal}
  {\bibinfo  {journal} {Surface Science}\ }\textbf {\bibinfo {volume} {387}},\
  \bibinfo {pages} {152 } (\bibinfo {year} {1997})}\BibitemShut {NoStop}%
\bibitem [{\citenamefont {Bondino}\ \emph {et~al.}(2002)\citenamefont
  {Bondino}, \citenamefont {Comelli}, \citenamefont {Baraldi}, \citenamefont
  {Rosei}, \citenamefont {Lizzit}, \citenamefont {Goldoni}, \citenamefont
  {Larciprete},\ and\ \citenamefont {Paolucci}}]{Bondino2002}%
  \BibitemOpen
  \bibfield  {author} {\bibinfo {author} {\bibfnamefont {F.}~\bibnamefont
  {Bondino}}, \bibinfo {author} {\bibfnamefont {G.}~\bibnamefont {Comelli}},
  \bibinfo {author} {\bibfnamefont {A.}~\bibnamefont {Baraldi}}, \bibinfo
  {author} {\bibfnamefont {R.}~\bibnamefont {Rosei}}, \bibinfo {author}
  {\bibfnamefont {S.}~\bibnamefont {Lizzit}}, \bibinfo {author} {\bibfnamefont
  {A.}~\bibnamefont {Goldoni}}, \bibinfo {author} {\bibfnamefont
  {R.}~\bibnamefont {Larciprete}}, \ and\ \bibinfo {author} {\bibfnamefont
  {G.}~\bibnamefont {Paolucci}},\ }\href {\doibase 10.1103/PhysRevB.66.075402}
  {\bibfield  {journal} {\bibinfo  {journal} {Phys. Rev. B}\ }\textbf {\bibinfo
  {volume} {66}},\ \bibinfo {pages} {075402} (\bibinfo {year}
  {2002})}\BibitemShut {NoStop}%
\bibitem [{\citenamefont {de~Lima}\ \emph {et~al.}(2020)\citenamefont
  {de~Lima}, \citenamefont {Greber},\ and\ \citenamefont
  {Muntwiler}}]{Lima2020}%
  \BibitemOpen
  \bibfield  {author} {\bibinfo {author} {\bibfnamefont {L.~H.}\ \bibnamefont
  {de~Lima}}, \bibinfo {author} {\bibfnamefont {T.}~\bibnamefont {Greber}}, \
  and\ \bibinfo {author} {\bibfnamefont {M.}~\bibnamefont {Muntwiler}},\ }\href
  {\doibase 10.1088/2053-1583/ab81ae} {\bibfield  {journal} {\bibinfo
  {journal} {2D Materials}\ }\textbf {\bibinfo {volume} {7}},\ \bibinfo {pages}
  {035006} (\bibinfo {year} {2020})}\BibitemShut {NoStop}%
\bibitem [{\citenamefont {Pendry}(1980)}]{Pendry1980}%
  \BibitemOpen
  \bibfield  {author} {\bibinfo {author} {\bibfnamefont {J.~B.}\ \bibnamefont
  {Pendry}},\ }\href {\doibase 10.1088/0022-3719/13/5/024} {\bibfield
  {journal} {\bibinfo  {journal} {Journal of Physics C: Solid State Physics}\
  }\textbf {\bibinfo {volume} {13}},\ \bibinfo {pages} {937} (\bibinfo {year}
  {1980})}\BibitemShut {NoStop}%
\end{thebibliography}%
\end{document}


\title{Supplemental Material for  Selecting ``Convenient Observers" to Probe the Atomic Structure of Epitaxial Graphene Grown on Ir(111) via Photoelectron Diffraction.}

\author{Lucas Barreto}
\affiliation{Centro de Ci\^{e}ncias Naturais e Humanas, Universidade Federal do ABC, Santo André 09210-580, SP, Brazil}

\author{Luis Henrique de Lima}
\affiliation{Centro de Ci\^{e}ncias Naturais e Humanas, Universidade Federal do ABC, Santo André 09210-580, SP, Brazil}

\author{Daniel Coutinho Martins}
\affiliation{Centro de Ci\^{e}ncias Naturais e Humanas, Universidade Federal do ABC, Santo André 09210-580, SP, Brazil}

\author{Caio Silva}
\affiliation{Instituto de F\'{i}sica Gleb Wataghin, Universidade Estadual de Campinas, Campinas 13083-859, SP, Brazil}

\author{Rodrigo Cezar de Campos Ferreira}
\affiliation{Instituto de F\'{i}sica Gleb Wataghin, Universidade Estadual de Campinas, Campinas 13083-859, SP, Brazil}

\author{Richard Landers}
\affiliation{Instituto de F\'{i}sica Gleb Wataghin, Universidade Estadual de Campinas, Campinas 13083-859, SP, Brazil}

\author{Abner de Siervo}
\affiliation{Instituto de F\'{i}sica Gleb Wataghin, Universidade Estadual de Campinas, Campinas 13083-859, SP, Brazil}

\maketitle
\section{XPD experimental details}
\sloppy{
The XPD measurements were performed at the PGM beamline of the Brazilian 
Synchrotron Light Source \cite{Cezar2013}.  We excited the Ir 4f core-level with linearly polarized \mbox{122 eV} photons parallel to the plane of the Synchrotron ring. Using an omicron HA125HR hemispherical electron analyzer with multiple channeltron detection, we recorded the photoemission intensities as a function of the polar ($\theta$) and the azimuthal ($\phi$) angles, where the polar angle $\theta=0$ means normal emission mounted with the energy dispersion in the same plane of the synchrotron ring. In total, we collected 1680 spectra  varying $\theta$ from 19$^{\circ}$ to 79$^{\circ}$, and $\phi$ over a 240$^{\circ}$ range, in steps of {3$^{\circ}$} for both angles. XPD experiments were performed in a low-magnification mode, which, combined with the set entrance and exit slits of the analyzer, results in an angular resolution of  2$^{o}$.  The detector was fixed, and the sample was rotated by a homemade manipulator. The angle between the beamline and the detector was 60$^{\circ}$.} 

\begin{figure*}[!htb]
\centering
\includegraphics[scale=0.53]{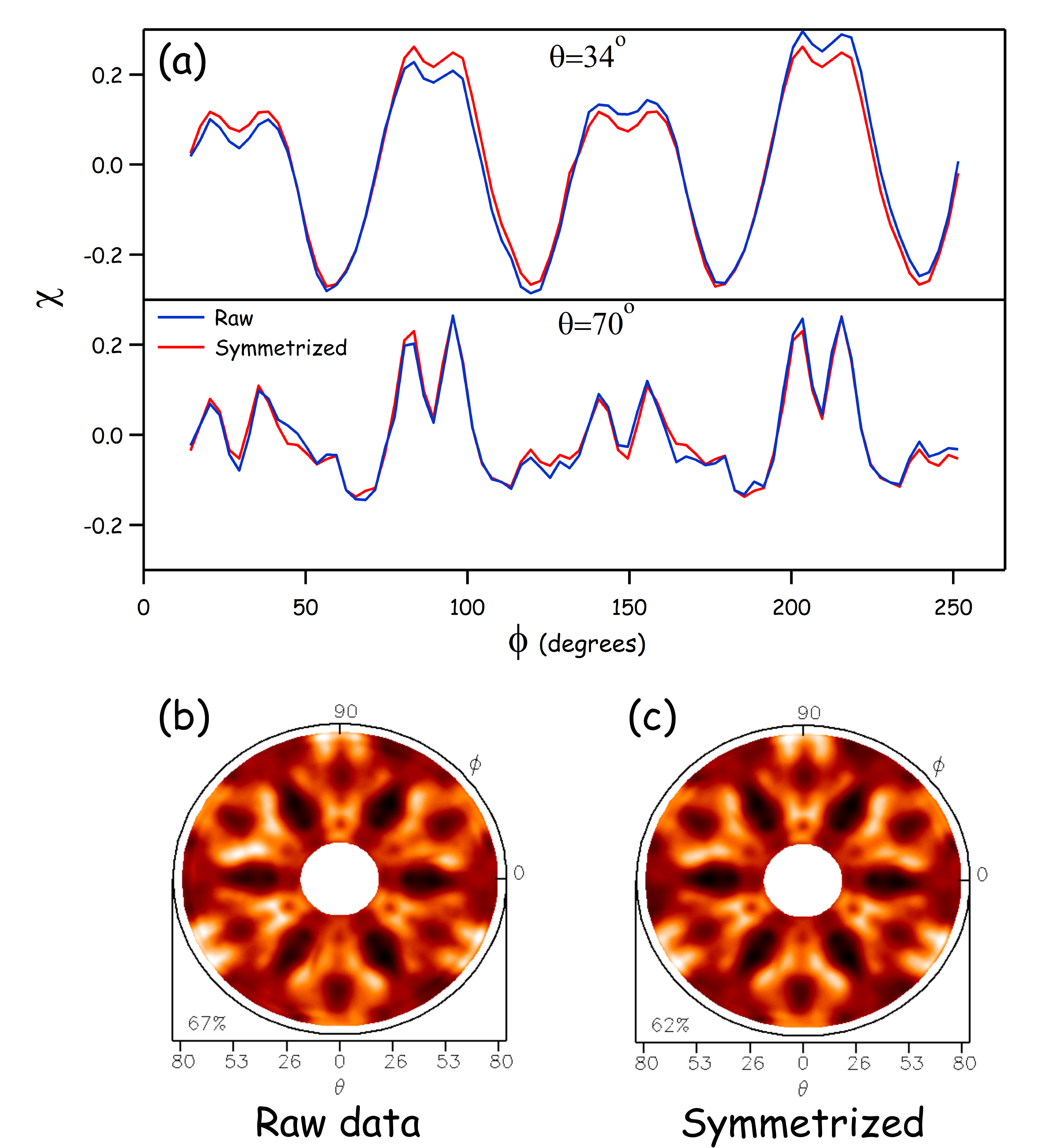}
\caption{(a) Anisotropy as a function of $\phi$ angle for $\theta=34^{\circ}$ (top) and $\theta=70^{\circ}$ (bottom). The $\chi$ function calculated from equation \ref{chi} is shown in blue. The symmetrized data exhibited in red is obtained after a Fourier analysis of the blue curve. (b) and (c) exhibit the XPD patterns for the raw and symmetrized data respectively.}
\label{fourier}
\end{figure*}

The spectra were fitted by Gaussian curves, and the photoemission intensity anisotropy function was determined by:

\begin{equation}\label{chi}    
\chi(\theta,\phi)=\frac{I(\theta,\phi)-I_{o}(\theta,\phi)}{I_{o}(\theta,\phi)}
\end{equation}
\noindent where $I(\theta,\phi)$ is the peak intensity, and $I_{o}(\theta,\phi)$ computes the mean intensity for each $\theta$ taking into account the synchrotron beam intensity decay. Particularly, we assume that $I_{o}(\theta,\phi)$ is a second-order polynomial. Due to the problem's symmetry, we applied a Fourier filter to extract the periodic component of $\chi$. Figure \ref{fourier}a exhibits the anisotropy of the Ir SS peak before and after the Fourier symmetrization for two $\theta$ values. Furthermore, their corresponding XPD patterns are shown in figures \ref{fourier}b-c. Since we only measured the $\phi$ angle over a 240$^{\circ}$ range, all the XPD pattern figures shown in this work are constructed by repeating the first 120$^{\circ}$ degrees. However, for the analysis, we only considered the measured dataset. 

\section{Multiple Scattering Calculations}

The multiple scattering calculations were performed using the MSCD computational package, which is based on the Rehr and Albers (RA) formalism \cite{Rehr1990,Chen1998}. Particularly, in the presented results, we considered up to eight multiple scattering events and fourth-order RA expansion. Furthermore, we used the standard configuration of the MSCD package in which the final state ($\ell_f$) is given by $\ell_f=\ell_i \pm 1$, where $\ell_i$ is the orbital angular momentum quantum number of the initial state \cite{Chen1998}. In this work, $\ell_i =3$.  
    
\section{Structure optimization}
The comparison between the theoretical and the experimental data is done through the reliability factor $R_a$\cite{Siervo2002, Soares2002} :
 \begin{equation}\label{Ra}    
R_a=\sum_i\frac{(\chi^{i}_{t}-\chi^{i}_{e})^2}{(\chi^{i}_{t})^2+(\chi^{i}_{e})^2}
\end{equation}

\noindent where the indexes $t$ and $e$ indicate, respectively, theoretical and experimental values, and the sum is performed over the entire measured dataset. It is important to stress that the surface structure optimization and the R factor calculation were performed considering the measured data; the replicated data set is used only for better visualization of the XPD pattern.  

\begin{figure*}[!htb]
\centering
\includegraphics[scale=0.4]{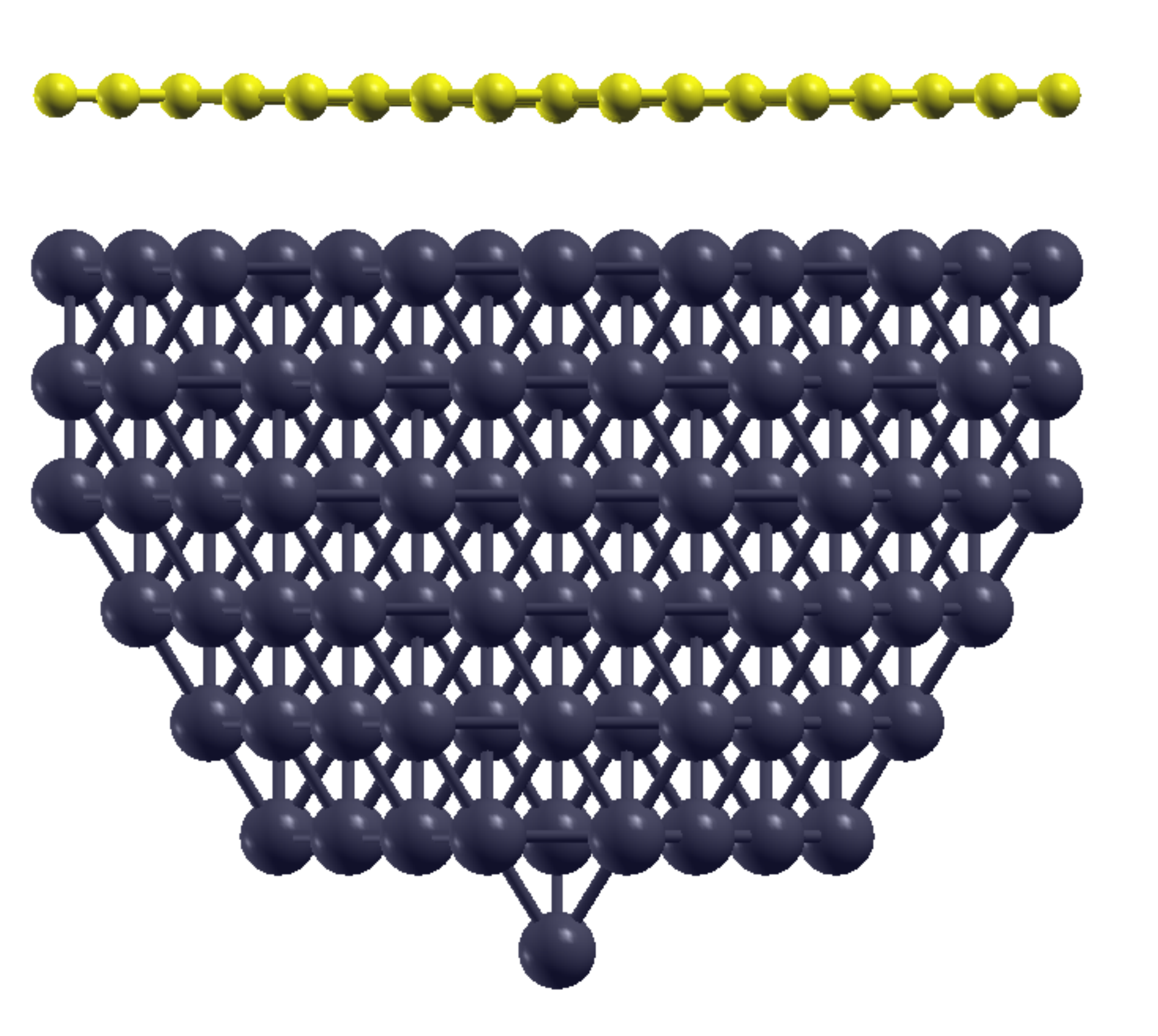}
\caption{Side view of the fcc cluster. Carbon and Iridium atoms shown in yellow and grey respectively. }
\label{cluster}
\end{figure*}

The theoretical modeling is based on a multiple scattering cluster calculation. In our case, the cluster presents a semi ellipsoidal shape with radius and depth equal to  11 {\AA}  and \mbox{17 {\AA}}, respectively, thus it has six Iridium layers and the graphene sheet as shown in figure \ref{cluster}. Furthermore, we assumed that the graphene profile on iridium is given by a distance offset ($z_o$) plus a superposition of the three Gaussians characterized by two parameters related to the amplitude and the width $(A_{i}, B_{i})$. 

The structural parameters were optimized using grids. Basically, we varied $z_o$ using predefined steps ($0.02$ \AA), and, for each $z_o$, we performed two-dimensional grids varying $A_{i}$ and  $B_{i}$ for the selected emitters. Then, we summed the three emitters XPD patterns incoherently, searching for the minimum $R_a$ of the combination. Roughly, we evaluated hundreds of combinations.  Independently, non-structural parameters were optimized. Specifically, we found the inner potential and the substrate surface Debye temperature to be equal to $8.0 \pm 1.0  $ eV and $180 \pm 20 $ K, respectively.

\section{Error analysis} \label{error}
As already mentioned, in the XPD structural determination, one uses a set of parameters $x_{i}$ to calculate the function $\chi^{i}_{t}$ and then compares it with the experimental data set through a reliability factor. In practice, a set of physical quantities is varied until the best agreement between theory and experiment is achieved. In this context, let $\vec{x}=(x_1, x_2, ..., x_i,...,x_n)$ be a vector in which each component represents a parameter to be optimized. Therefore, examining equation \ref{Ra}, it is desirable to find a vector $\overrightarrow{x^{min}}=(x^{min}_1, x^{min}_2, ..., x^{min}_i,...,x^{min}_n)$ that minimizes the $R_a$ value ($R^{min}_a$).   

A displacement in a parameter $x_i$ must lead to a variation in the calculated XPD pattern and, consequently, in the $R_a$ value. If $R_a$ is independent of $x_{i}$, it indicates that the technique is not sensitive to $x_i$,  or the modeling methodology is inappropriate to describe the experiment. Inversely, if there is a strong dependence between $R_a$ and $x_{i}$,  the experiment resolves  $x_{i}$ properly. Therefore, we can evaluate the precision of determining $x_{i}$ by its influence on $R_a$. 

Based on this premise, suppose that the vector $\overrightarrow{x^{min}}$ is known, and we want to determine the error of the parameter $x^{min}_i$. In this case, we calculate $R_a$ for different $x_i$ values in the vicinity of $x^{min}_i$, retaining the other variables at the minimum. In the proximity of the minimum, we can approximate $R_a$($x_i$) by: 

\begin{equation}\label{Ramin}    
R_a(x_i)=R^{min}_a+C(x_{i}-x^{min}_i)^2
\end{equation}  

\noindent where $C$ is the polynomial curvature. From the experimental point of view,  $C$ is a scaling factor that relates a variation on $R^{min}_a$ with a displacement on $x^{min}_i$. This connection provides a quantitative procedure to estimate the $x^{min}_i$ error. So, we can rewrite equation \ref{Ramin} as:  

\begin{equation}\label{Ramin2}    
x_{i}-x^{min}_i=\sqrt{\frac{[R_a(x_i)-R^{min}_a]}{C}} \Rightarrow \Delta x_i=\sqrt{\frac{\Delta R_a}{C}}.
\end{equation}            
 
\begin{figure*}[!htb]
\centering
\includegraphics[scale=0.70]{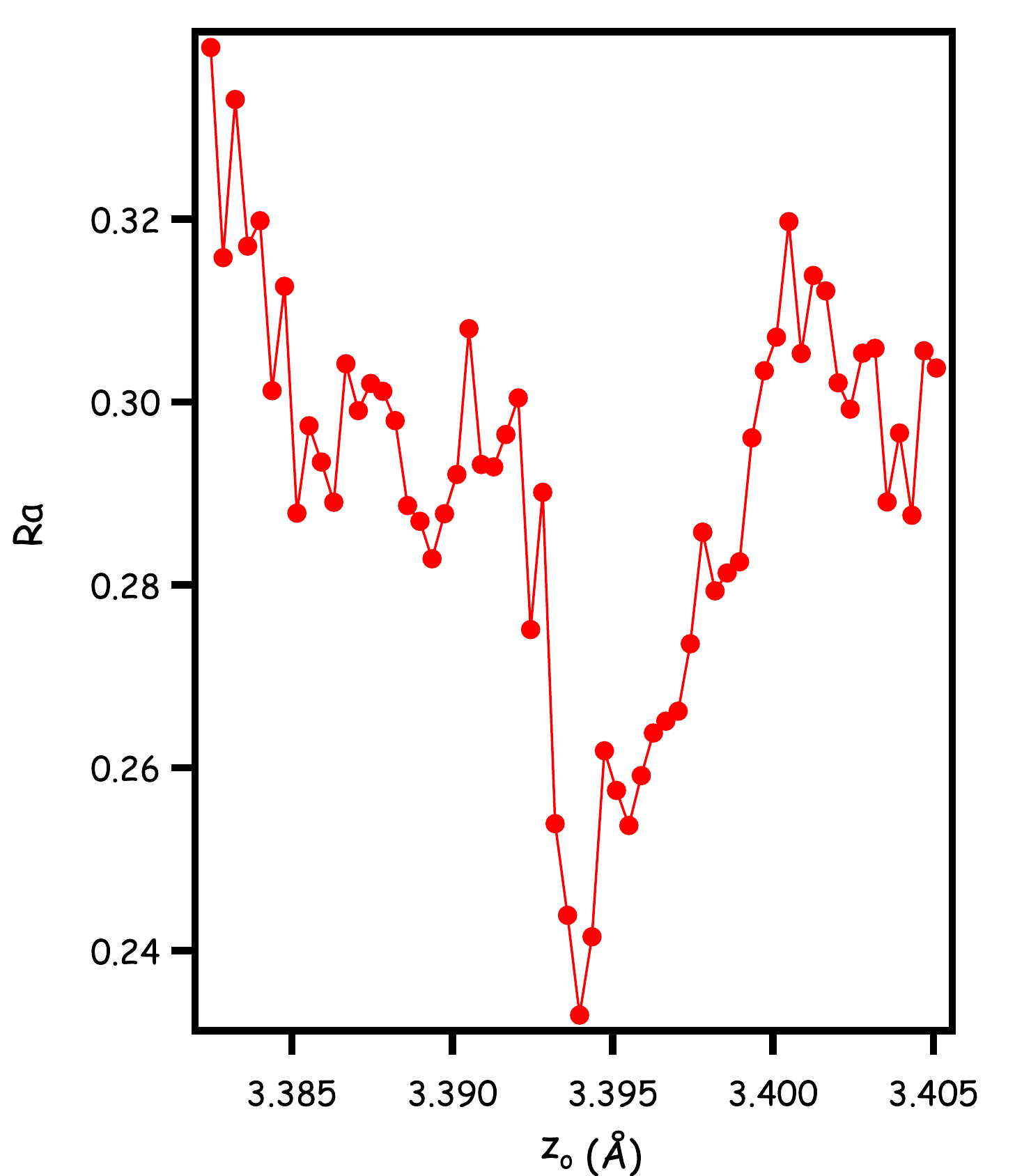}
\caption{$R_a$ as a function of $z_o$. In the vicinity of the well, we approximate the curve  by equation \ref{Ramin2} to determine the error.}
\label{Razo}
\end{figure*}

From equation \ref{Ramin2}, we state that the error of the parameter $x^{min}_i$ is equal to $\Delta x_i $. Finally, it is necessary to establish a procedure to determine $\Delta R_a$ based on the dataset. As previously suggested \cite{Booth1997, Bondino2002, Lima2020}, we extend the method used in LEED proposed by J. B. Pendry \cite{Pendry1980}.  For a set of N distinguishable peaks, the statistical variance of the reliability factor satisfies \mbox{$\Delta R= R\sqrt{2/N}$}. For the data presented in the main text, $N \sim 86$. As an example, figure \ref{Razo} shows the dependence of $R_a$ with z$_o$ from which we estimate the parameter  $C$ and then the error. The same approach is used to determine all the error bars listed on table I in the main text.             
 
\FloatBarrier 
\section{Influence of the Graphene Sheet on the XPD Pattern.}
\begin{figure*}[!htb]
\centering
\includegraphics[scale=0.27]{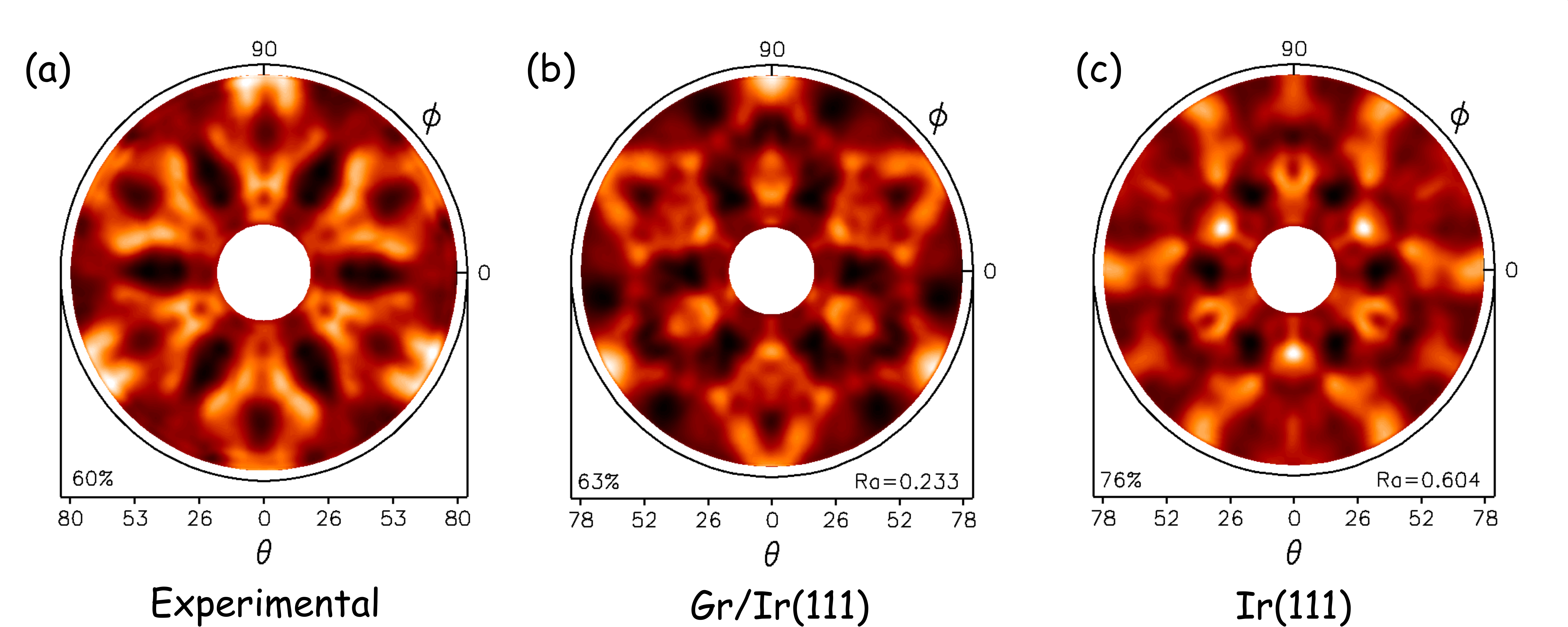}
\caption{(a) Experimental and (b)-(c) theoretical XPD patterns. (b) and (c) are calculated for Gr/Ir(111) and clean Ir(111), respectively. }
\label{Irclean}
\end{figure*}

As discussed in section \ref{error}, one can evaluate the sensitivity of the experiment by modifying a parameter in the modeling process and examine how it changes the experimental description. Based on this, we calculated the Ir4f SS XPD pattern of a cluster without the graphene layer to verify that the carbon atoms produce a measurable signal. Examining figure \ref{Irclean}, it is possible to compare the experimental (figure \ref{Irclean}a) and theoretical XPD patterns (figures \ref{Irclean}b-c), where a clear visual difference between the theoretical XPD patterns for Gr/Ir(111) and Ir(111) is observable. Additionally, a comparison with experimental data results in $Ra=0.60$ for clean Ir(111), which is higher compared to the model described in the main text, including the individual chosen sites. Those results strongly suggest that the presence of the graphene modifies the XPD Ir 4f SS anisotropy.

\FloatBarrier
\bibliography{subref}